\documentclass[fleqn,10pt]{wlscirep}
\usepackage[utf8]{inputenc}
\usepackage[T1]{fontenc}
\usepackage{graphicx}
\usepackage{dcolumn}
\usepackage{csquotes}
\usepackage[capitalise,nameinlink]{cleveref}
\usepackage{gensymb}
\usepackage[justification=raggedright]{caption}
\usepackage{subcaption}
\usepackage[version=4]{mhchem}
\usepackage{float}

\title{On the detection of alpha emission from a low-voltage DC deuterium discharge with palladium electrodes}

\author[1,*]{Erik P. Ziehm}
\author[1]{George H. Miley}
\affil[1]{Department of Nuclear, Plasma, and Radiological Engineering, University of Illinois Urbana-Champaign, Urbana, 61801, USA}

\affil[*]{ziehm1@illinois.edu}

\begin{abstract}
The possibility of nuclear reactions within a low voltage DC deuterium discharge with palladium electrodes is addressed. The solid-state nuclear track detector known as CR-39 was chosen to investigate the emission of energetic charged particles from the electrodes. A partially automated imaging platform and feature classification process was developed to scan the CR-39 surface and detect tracks. Typical discharge parameters were 10 Torr deuterium, 5-7 mm electrode gap distance, 20-40 mA/cm$^2$ current density, and -500 $\pm$ 100 V cathode bias. After discharge treatments to varied ion-cathode fluences, tracks formed in CR-39 which consistently corresponded to 138 $\pm$ 21 keV alpha particles emitted from the palladium electrodes. The track densities for deuterium discharges were often $\sim$100 times above controls with hydrogen and helium. Currently, there are no known mechanisms to accelerate ions to these energies within the apparatus. The production of energetic alpha particles with no source of helium or a means to accelerate the ions to such high energies indicate a nuclear origin. From particle trajectory estimates based on track geometries, it was concluded the reactions originated at the Pd electrodes and not from external sources such as atmospheric radon or cosmic rays. 

\end{abstract}
\begin{document}

\flushbottom
\maketitle

\thispagestyle{empty}

\section*{Introduction}\label{sec:intro}

This work undertook a systematic study of unexplained energetic charged particle emission in deuterium-metal experiments\cite{dong1991, Chambers1991aers1991Searchlattices, Cecil1990, Cecil1991b, Lipson2005a, Taniguchi1989}. For example, Chambers et al.\cite{Chambers1991aers1991Searchlattices} observed $\sim$5 MeV particle emission during 350-400 eV deuteron bombardment of titanium or the work done by Cecil et al.\cite{Cecil1991b} showing the production of 1-10 MeV particles as a DC current was applied to titanium foils with high concentrations of deuterium. Neither group reached a firm conclusion on particle type. Chambers et al.'s results suggested protons, deuterons, or tritons while Cecil et al. suggested $^3$He or alphas. As these experiments are derived from interest in \textquote{cold fusion} within a heavy water electrolysis cell with a palladium cathode and platinum anode \cite{Fleischmann1989}, it is natural to focus on possible deuteron-deuteron fusion reaction enhancements due to internuclear many-particle processes in metal-hydrides leading to Coulomb barrier suppression\cite{Ichimaru1993}. However, the 1-10 MeV particle energies do not correspond to the nuclear byproducts associated with $dd$ reaction pathways \cite{rolfs1988cauldrons} and require an alternate explanation. 

Once laboratories began reporting other peculiar aspects on deuterium loading of metals, it was realized $dd$ fusion was too myopic of a view and the field took on a new name, Low Energy Nuclear Reactions (LENR). Miley et al. found another surprising feature of LENR wherein Neutron Activation Analysis (NAA) and Secondary Ion Mass Spectrometry (SIMS) showed transmutation reactions occur within thin films of nickel\cite{mileyQuantitativeObservationTransmutation1996} and titanium\cite{mileyUseCombinedNAA2005} coatings on microspheres in a packed-bed electrolytic cell. A particularly relevant hypothesis was the necessity for dense deuteron formations within the electrode lattice to produce LENR effects \cite{Miley2009}. 

The focus of the present study was then to create dense deuteron formations using ion bombardment of palladium electrodes within a DC discharge while monitoring for charged particle nuclear byproducts. It is not intended to provide a comprehensive history of LENR nor is it attempting to build a theoretical framework describing LENR phenomena. The reader is cautioned with the author's opinion that it is premature to propose a physical mechanism that governs the phenomena as the relevant parameters of the plasma-material interaction have yet to be studied, such as ion and excited species concentrations and energy distributions at the electrode surface along with possible long-lived metastable states. Further detail on the context of the research is provided in the author's Ph.D. thesis\cite{ziehmEXPERIMENTALINVESTIGATIONLOW2022}. 

\begin{figure}
    \centering
    \includegraphics[width=0.4\textwidth]{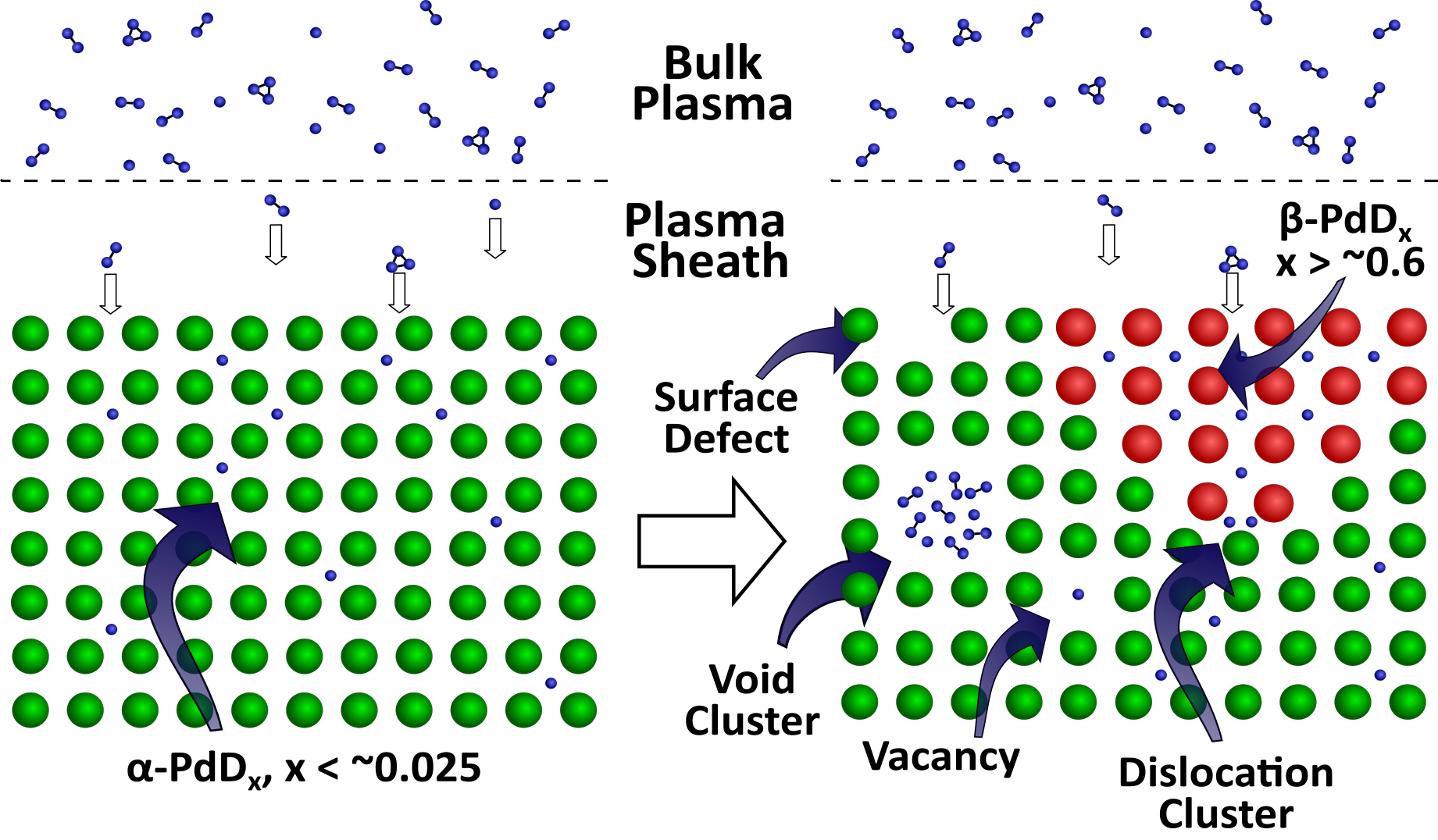}
    \caption{\label{fig: schematic of ion bombardment} Plasma ion implantation and defect formation. (Left) interstitial deuterons form a dilute phase metal-hydride, $\alpha$-PdD$_x$, with x $<$ $\sim$0.025. Where x is the atomic ratio [D]/[Pd]. (Right) high concentration phase formed in red, $\beta$-PdD$_x$ with x $>$ $\sim$0.6, and deuterium trapped at defects, e.g., vacancies, dislocations, and voids to form clusters.}
\end{figure}

It is important to note that the aforementioned studies of metal-hydride systems involve both the metal properties and the dynamics of how the hydrogen enters the metal. In the case of the present work utilizing a DC discharge, an interplay exists between incident plasma ions on the electrodes, defect formation, and the mobility and trapping of solute ions in the metal lattice. \cref{fig: schematic of ion bombardment} is a diagram that depicts the ion bombardment of a crystalline lattice. The left portion of \cref{fig: schematic of ion bombardment} is a palladium-deuteride composed of dilute deuteron interstitials. Upon continued deuteron implantation, various types of defect traps are formed along with a high concentration deuteride phase designated in red.

Multiple factors were taken into consideration to determine the optimal charged particle detection method. A time-integrative high detection efficiency feature was desirable to accumulate counts in the case of low emission rates in order to increase the signal-to-noise ratio. The detector must be unaffected by electromagnetic interference due to unsteady plasma characteristics which was a significant obstacle for Chambers et al.\cite{Chambers1991aers1991Searchlattices} using Silicon Surface Barrier Detectors (SSBD) and later also found by the author of this work. The detector must also be insensitive to visible to UV photons from discharges. Large detection areas with high spatial resolution were required to investigate the possibility of detecting localized \textquote{hot spots} of emission from distinct deuteron-clusters in the metal electrode. Lastly, the detector must be able to withstand high-vacuum and a low pressure hydrogen atmosphere. The Solid-State Nuclear Track Detector (SSNTD) commercially known as CR-39 was then chosen as the charged particle detector.

\begin{figure}
\centering
\includegraphics[width=0.4\textwidth]{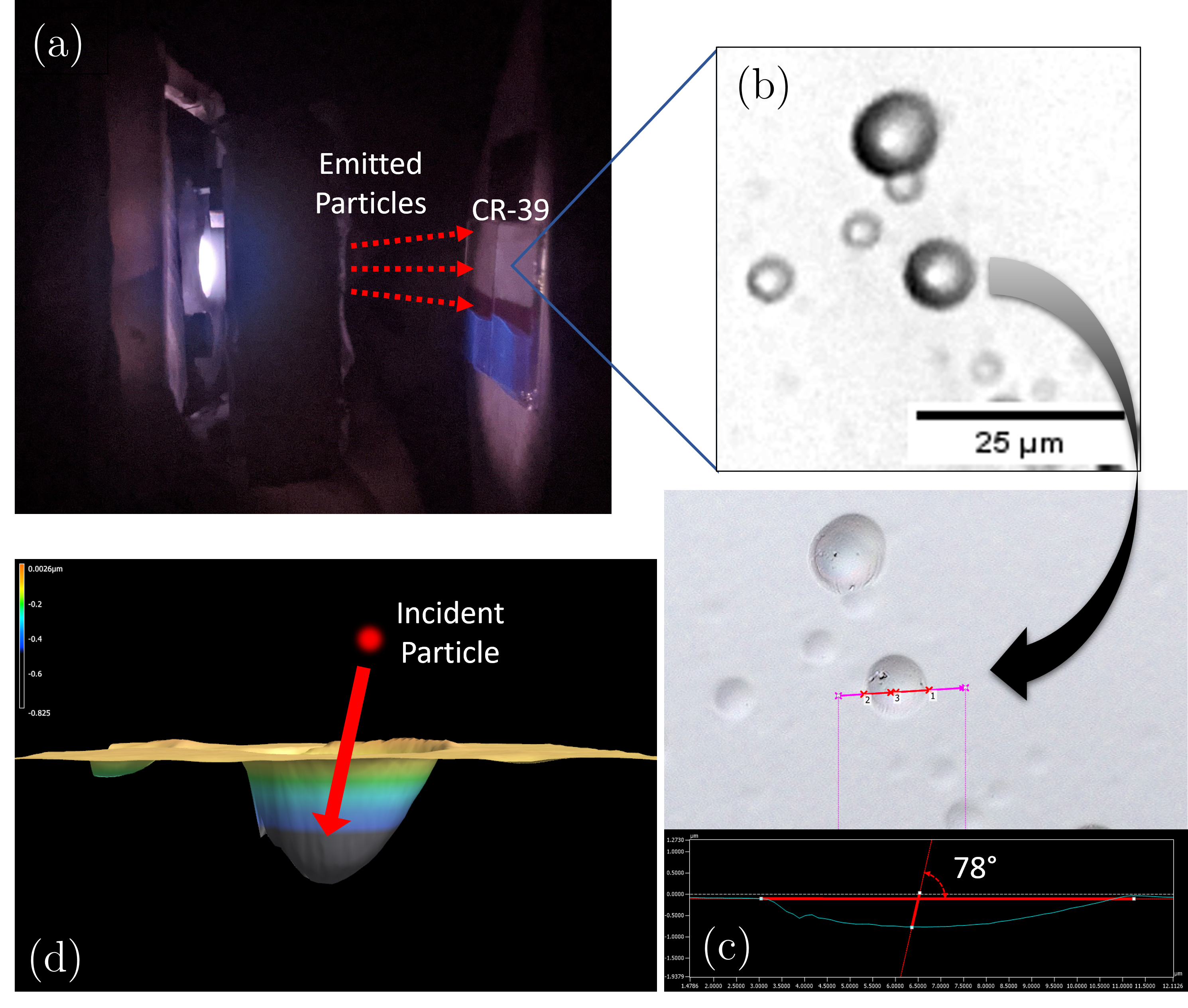}
\caption{(a) DC discharge with trajectories of emitted particles and CR-39. (b) optical image of tracks in CR-39. (c) the depth profile of a track along with an estimated incident angle. (d) 3D image of the track with a representation of the incident particle.}
\label{fig:off-normal tracks_1}
\end{figure}

\cref{fig:off-normal tracks_1} portrays the experimental approach of this research: tracks in CR-39 after a deuterium DC discharge with Pd electrodes that correspond to roughly 100 keV alpha particles. \cref{fig:off-normal tracks_1}a is an image of the discharge in 10 Torr of D$_2$ operating at 40 mA/cm$^2$ current density. \cref{fig:off-normal tracks_1}b is an optical image of the tracks using a reflected light microscope. \cref{fig:off-normal tracks_1}c shows the depth profile of a track using a 3D laser scanning confocal microscope, Keyence VK-X1000. Then \cref{fig:off-normal tracks_1}d is the 3D image of the track, along with a representation of the incident particle. A new automated CR-39 analysis technique amassed considerably more track data than previous studies, e.g., $\sim$10$^{3}$tracks/cm$^2$/day for a 1 cm$^2$ chip by Lipson et al.\cite{Lipson2005h} compared to $\sim$10$^{6}$tracks/cm$^2$/day for a 6.25 cm$^2$ chip in the current work. As helium was not present in the system and the energies of the alphas exceeded known acceleration mechanisms in the device, it was concluded that a nuclear reaction produced the particles. Other causes of tracks were investigated, such as charge build-up on the CR-39 surface but were inconsistent with the observed data. 

\section*{Methods}

\begin{figure}
    \centering
    \begin{subfigure}{0.45\textwidth}
        \includegraphics[width=1\linewidth]{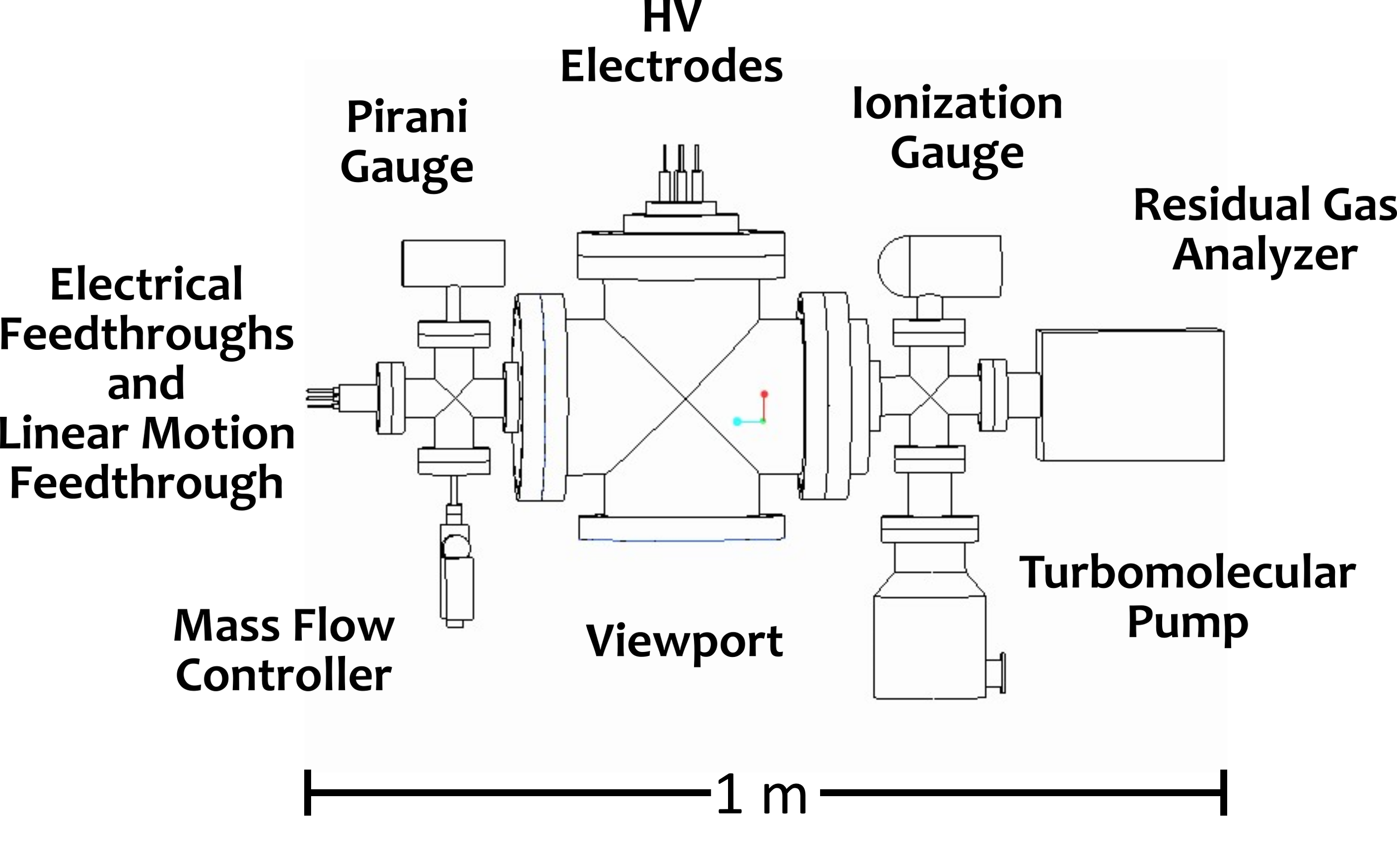}
        \caption{}
    \end{subfigure}
    \begin{subfigure}{0.45\textwidth}
    \includegraphics[width=1\linewidth]{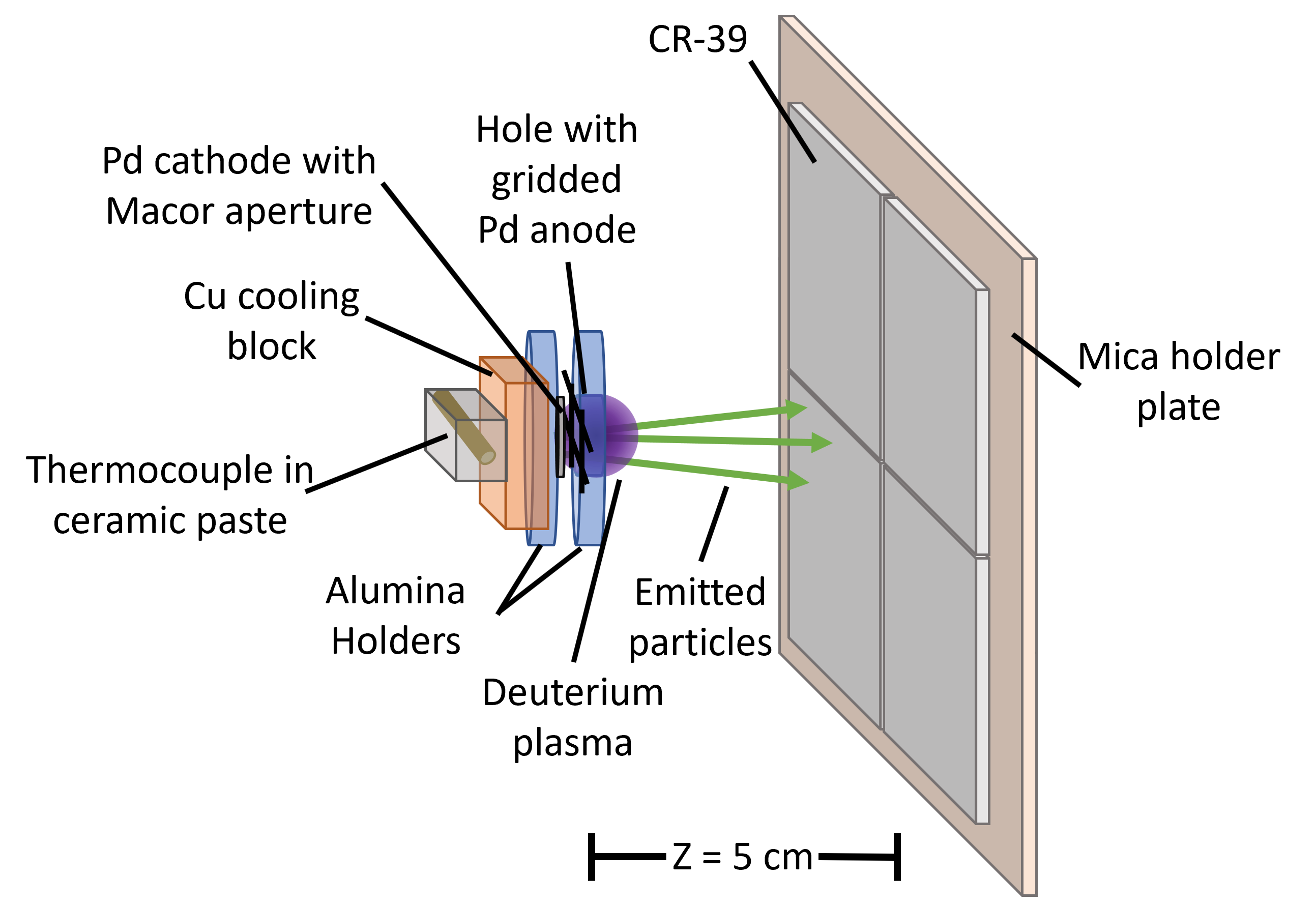}
    \caption{}
    \end{subfigure}
    \caption{(a) Schematic of vacuum system. (b) Electrode and CR-39 diagram.}
    \label{fig:MS - chamber diagram}
\end{figure}

A schematic of the vacuum system and electrode apparatus are shown in \cref{fig:MS - chamber diagram}. The chamber could reach below 10$^{-7}$ Torr. The main chamber is an 8-inch diameter cross wrapped in heating tape. Bakeouts were done for roughly 12 hours with the temperature set to 325$\degree$C at the center of the main chamber's external wall using a K-type thermocouple. The Residual Gas Analyzer (RGA) was used to verify adsorbed gases were released, such as water partial pressure below 1 $\times$ 10$^{-9}$ Torr. A thermal camera showed uneven temperature profiles of the chamber walls and was the cause of setting the set-point to 325$\degree$C to ensure the entire 8-inch diameter cross reached 200$\degree$C. The gases were purchased from Airgas, An Air Liquide Company. The purity for the hydrogen and helium gases was $99.999 \%$ while deuterium was $99.9999 \%$. The palladium electrodes were biased by a Glassman High Voltage EK series 600 W DC power supply. A LabView program was written to acquire the voltage, current, pressure, and an estimate for the cathode's ion fluence in 0.1 ms intervals. The fluence assumes the plasma is localized on the cathode, which was visually verified and did not take into account backscatter ions or secondary electron emission. 

The electrode discharge gap in \cref{fig:MS - chamber diagram}(b) was 5 mm while the active cathode area was 0.5 cm$^2$. One cause of concern for accurate modeling of the discharge is the anode grid sometimes deformed and warped which leads to an uneven electrode gap distance. A 1/8" thick Macor ceramic sheet with a 0.5 cm diameter circular window was placed over the cathode tab to prevent the discharge from interacting with the high voltage connection. Along with the Macor aperture, the electrode setup had the Pd electrodes inside solid alumina holders for electrical insulation. The cathode holder was joined to a copper cooling block using high-temperature silver paste. Tap water was used as the coolant at a flow of 3 L/min. The cathode was formed from palladium bullion from Valcambi Suisse with a purity of 99.95$\%$ which was rolled to 0.3 mm thick 1 cm x 0.7 cm tabs. Once the electrode holder was fixed in the system and baked out, the chamber was pressurized with the gas used for the test (hydrogen, deuterium, or helium) at 10 Torr and a discharge was struck under current control at 10 mA with the Pd tab positively biased, 300-400 V, for surface cleaning. The chamber was then pumped down again to high vacuum for 30 minutes to an hour and then the ion treatment began with the Pd tab biased negatively as the cathode with the grid anode grounded. 

The CR-39 holder was fixed to a linear motion feedthrough. To reduce CR-39 exposure to high temperatures during bakeouts, the holder was moved 13 cm away from the electrode at the edge of the main chamber where the 8-inch cross joins to a 2.75 inch cross which held the pirani gauge, electrical feedthroughs, and mass flow controller as shown in \cref{fig:MS - chamber diagram}(a). A thermocouple was placed on the CR-39 holder during a bakeout and it reached a maximum temperature of $95 \degree$C. The possibility of the bakeouts affecting the CR-39 and degrading the polymer was investigated by placing CR-39 inside the vacuum chamber in proximity to the external thermocouple which was held at $150 \degree$C for 12 hours. After cooling to room temperature and removal from the vacuum chamber, the CR-39 piece was irradiated with unattenuated \ce{^{241}Am} alphas. No noticeable effects were found on surface morphology and track diameters.

\subsection*{\label{subsec:intro cr39}Solid-State Nuclear Track Detectors}

SSNTDs are used as a passive radiation detector in a variety of fields, ranging from elementary particle physics, fission research, and atmospheric radon measurements to geology and biology. The poly-allyl diglycol carbonate polymer (C$_{12}$H$_{18}$O$_7$) known as CR-39 is the most popular commercial track detector. Its ease of sample preparation and low cost are two of the primary factors in its widespread use. Other advantages are its light weight, lack of external electronic systems susceptible to noise, ability to be cut and placed in locations and environments which hinder the use of other detection methods, and its intrinsic detection efficiency nearing 100$\%$ for charged particles with normal incidence over a wide energy range. The method can be used in high or low dose radiation environments and can permanently save the incident particle’s trajectory with micron spatial resolution \cite{durraniSolidStateNuclear2013,fleischerNuclearTracksSolids1975}. Not only does it store the particle's trajectory, but it is also possible to deduce the particle charge, mass, and energy from the track characteristics given appropriate calibration standards.

\begin{figure}
    	\centering
	\includegraphics[width=0.6\columnwidth]{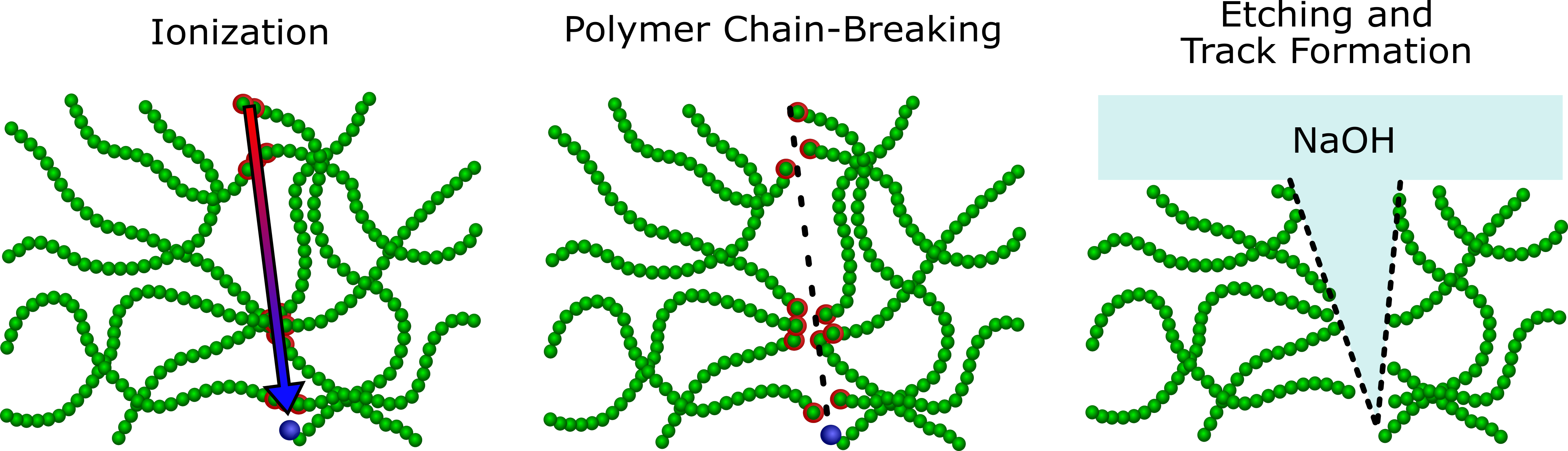}
    	\caption{\label{fig:CR-39 - stages of CR-39 development}Stages of charged particle detection in CR-39 polymer.}
\end{figure}

When an energetic particle passes through the polymer, it produces along its path a region that is more sensitive to chemical etching than the rest of the bulk. \cref{fig:CR-39 - stages of CR-39 development} depicts the stages of charged particle detection in polymer SSNTDs. The particle deposits energy primarily following the electronic stopping power in the substrate, which leaves an ionization trail. Bonds break in the polymer, and radicals and defects form. The open ends of the chains can then be easily attacked and preferentially etched. After treatment with an etching solution, e.g., NaOH or KOH, tracks remain as pits in the plastic. 

\begin{figure}
  \centering
      \begin{subfigure}{0.3\textwidth}
        \includegraphics[width=1\textwidth]{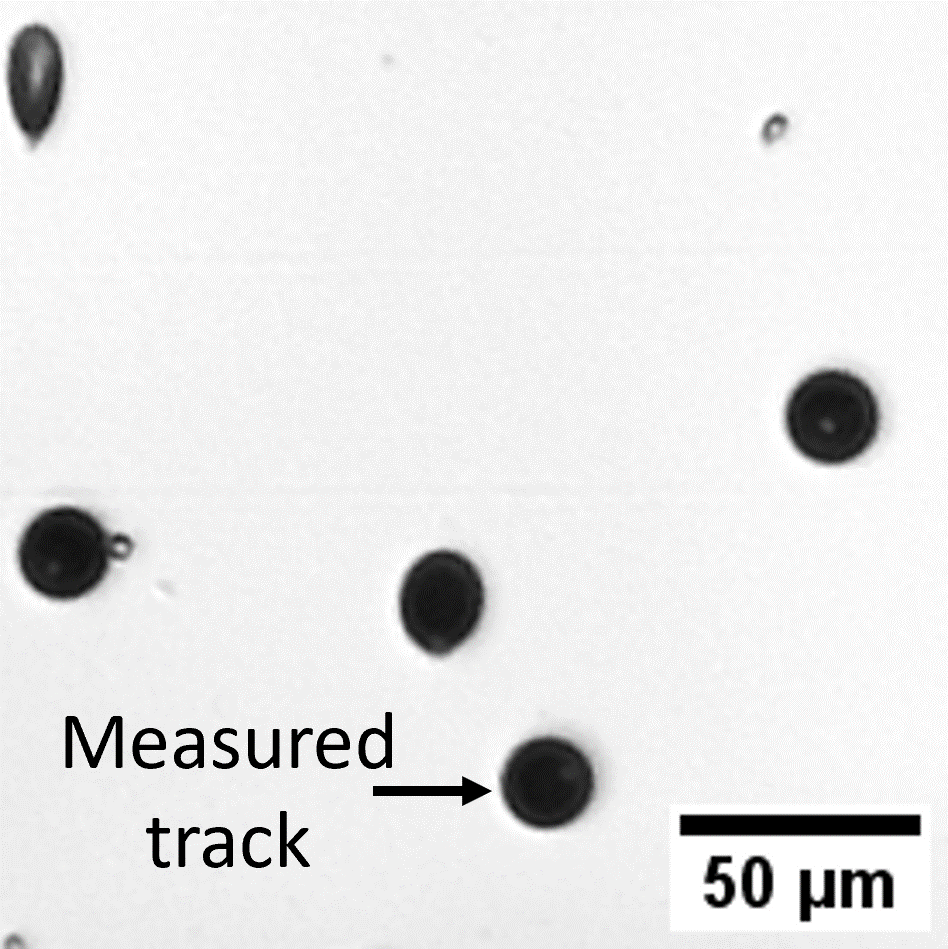} 
        \caption{}
      \end{subfigure}
      \begin{subfigure}{0.3\textwidth}
        \includegraphics[width=1\textwidth]{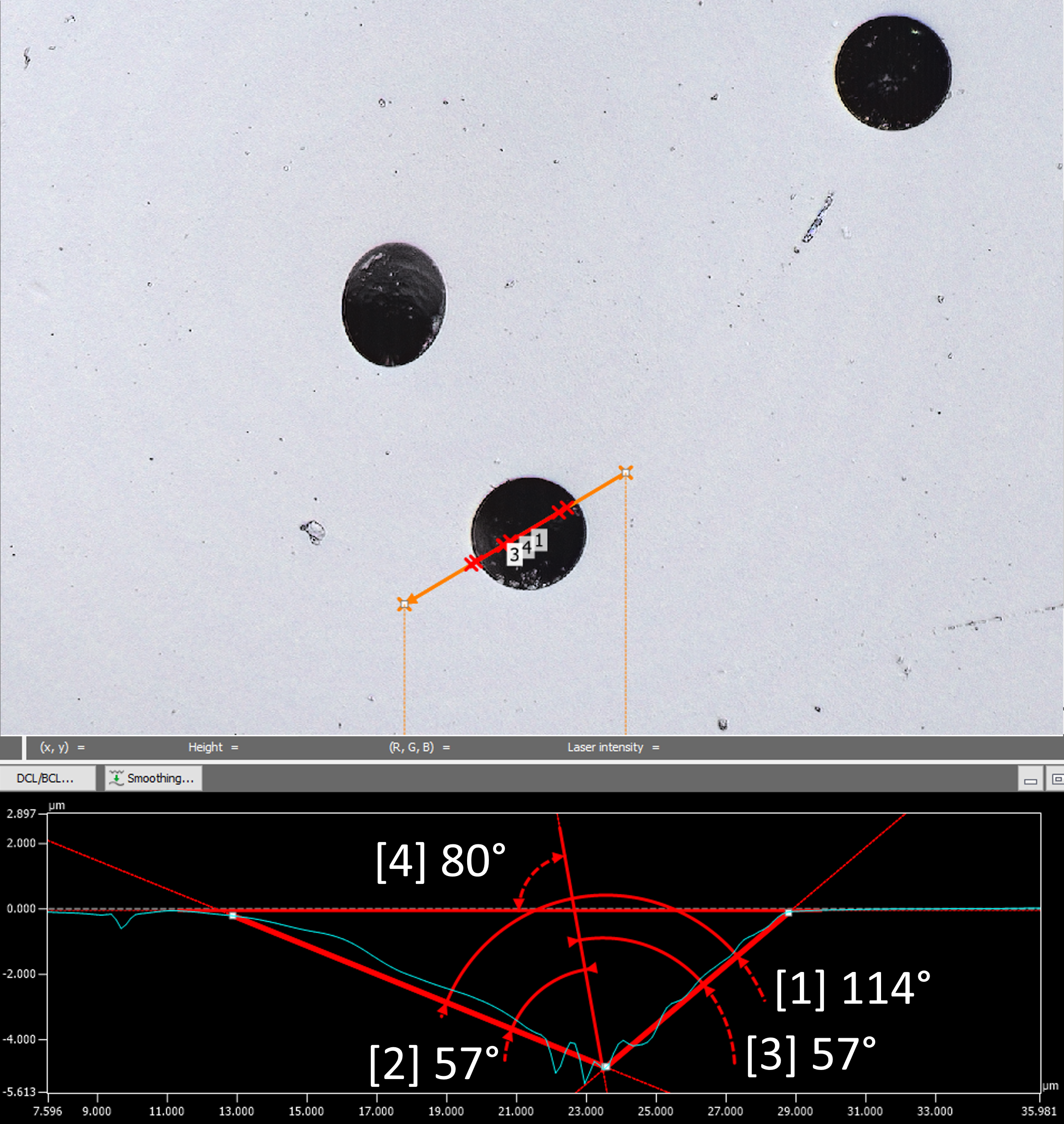}
        \caption{}
      \end{subfigure}
      \caption{Tracks produced from \ce{^{241}Am} source. Etched in 6.25 M NaOH for one hour at $98\pm1\degree$C. (a) optical image using a reflected light microscope. (b) 3D laser confocal microscope image showing depth profile of individual track.}
      \label{fig:CR-39 - example of tracks}
\end{figure}

The etched track morphology then provides information as to the type of charged particle, along with a rudimentary form of energy spectrometry. A calibration curve of the etched pit diameter versus incident energy is formed for particles of interest for each batch of CR-39 as the detector's response to etching varies based on manufacturing procedures. Most studies using CR-39 for energy spectrometry focus on normal incidence angles, which limits quantifiable particle tracks \cite{Seimetz2018a,khanLightChargedParticle1984,guoChapterSolidStateNuclear2012,Jeong2017,Sinenian2011a,zhangEnergyCalibrationCR392019}. An example of off-normal tracks from an unattenuated \ce{^{241}Am} source (dominant alpha population at 5.486 MeV) then etched in a 6.25 M solution of NaOH for one hour at $98 \pm 1 \degree$C is shown in \cref{fig:CR-39 - example of tracks}(a). One can see the tracks vary in shapes, sizes, and contrast. Tracks that are more circular at the surface can be discriminated, and then the features can be analyzed and correlated to particle type and energy. \cref{fig:CR-39 - example of tracks}b is obtained from a 3D laser confocal microscope. It shows the depth profiles used for angle of incidence estimates. The track of interest is annotated in \cref{fig:CR-39 - example of tracks}(a), and it is estimated to have an $80 \degree$ angle of incidence with respect to the surface of the CR-39, see angle [4] in \cref{fig:CR-39 - example of tracks}(b).

\subsection*{\label{subsec:intro semi-automated imaging system}Semi-Automated Imaging Procedure}

A novel imaging system was developed to automatically obtain images over large surface areas of CR-39. The apparatus is described in \cite{Ziehm2018c,ziehmAdvancementsSemiAutomatedCR392020,ziehmEXPERIMENTALINVESTIGATIONLOW2022}. The objective was to autonomously obtain images of an entire CR-39 sample over large surface areas, e.g., 6.25 cm$^2$. For that purpose, two components were necessary: an imaging system and a stage capable of moving the sample relative to the imaging optics. The sample movement was accomplished using an X-Y plotter stage with an arm extension connected to a microscope slide. A user-defined path, dependent on the sample dimensions, determines the sample movement, while the microscope captures images at an optimal focus based on a Laplacian image filter. The automated stage movement and focusing was implemented with a python script and an Arduino microcontroller. 

\begin{figure}
    	\centering
    	\includegraphics[width=0.7\columnwidth]{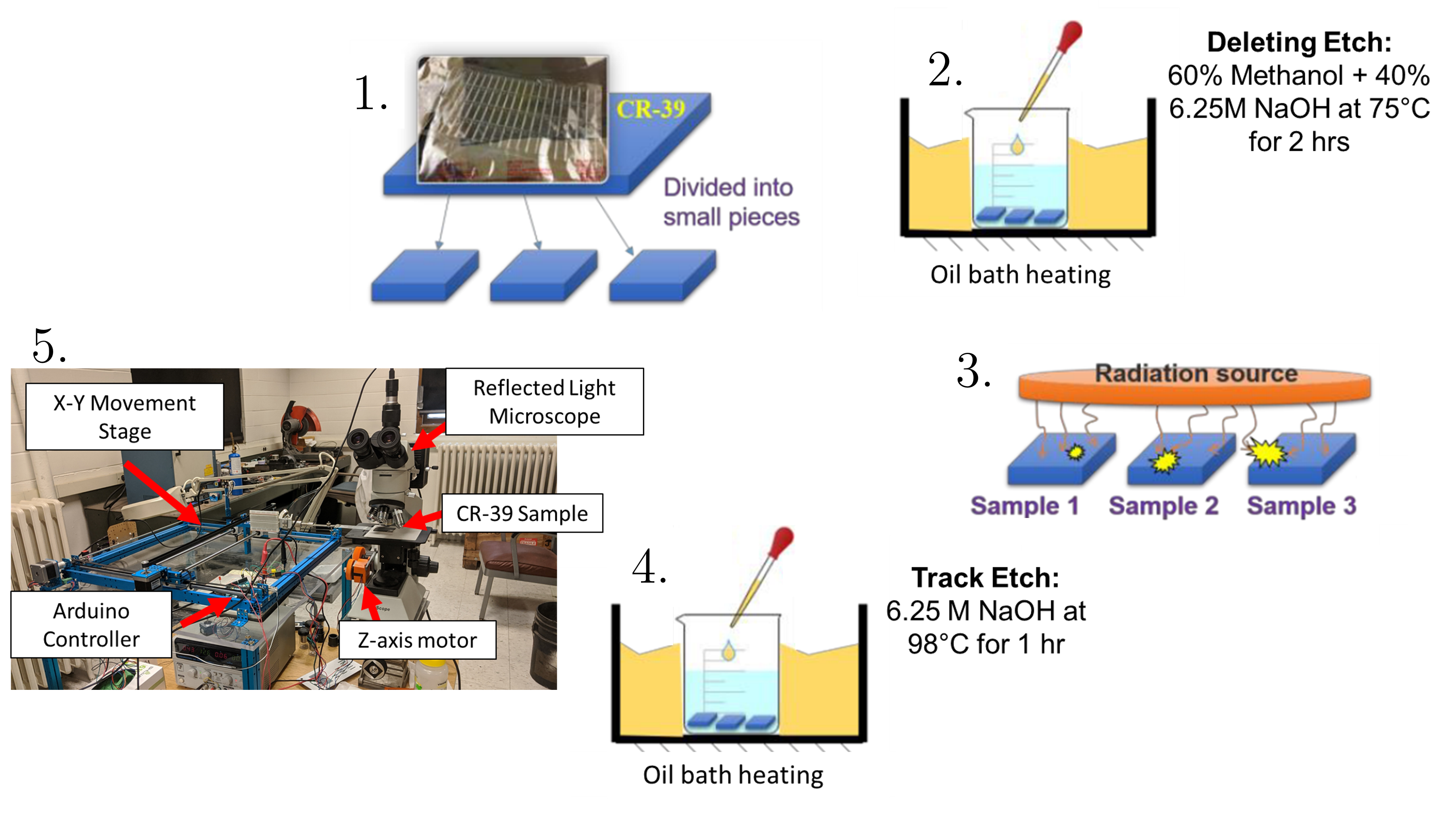}
    	\caption{\label{fig:CR-39 - example of CR-39 treatment}An example CR-39 experimental procedure.}
\end{figure}

Before any experiments, a deleting etch is done to remove background tracks and surface damage. The deleting etch ensures that any subsequent tracks found originated during the time between the deleting etch and the following track etch. The deleting etch solution is 60$\%$ methanol and 40$\%$ 6.25 M NaOH at $75 \pm 5 \degree$C for 2 hours, which removes the top $\sim$150 \textmu m of material. The methanol reduces the track-etch sensitivity and allows for an even etch profile. After the experiments, etching was done with 6.25 M NaOH at $98 \degree$C for one hour. An example of a CR-39 treatment procedure is shown in \cref{fig:CR-39 - example of CR-39 treatment} where, (1) the CR-39 is cut into desirable sizes using a laser cutter, (2) deleting-etch, (3) the samples are exposed to a radioactive source, (4) track-etching, and (5) the imaging system. 

Once a surface image of the sample is obtained, a machine learning pipeline was implemented to reduce human error while detecting particle tracks. The software used was $ilastik$ - an easy-to-use tool to perform segmentation and classification during image processing without substantial computational programming expertise \cite{Berg2019}. It includes supervised machine-learning techniques that rely on the user to classify objects to effectively \textquote{train} the software from examples such as tracks, defects, scratches, dust, etc. The training data comprised 58 example images of a sample irradiated with \ce{^{241}Am} along with samples that were left open to atmosphere for 7 days. The samples were then etched for 1 hour at $98 \pm 1 \degree$C in 6.25 M NaOH. 

The analysis pipeline was tested with samples that were irradiated with the \ce{^{241}Am} source to determine the detection efficiency (the percentage of tracks correctly identified in an image). The samples were first put through a deleting etch, irradiated with the \ce{^{241}Am} source, and etched again for tracks. Thirteen images with $\sim$800 tracks went through the analysis pipeline and the total standard error compared to manual counting was 2.28$\%$. For subsequent analysis, this error was increased to 15$\%$ to be conservative. 

\subsection*{\label{Sec: Am-241 calibration}Minor Axis Versus Alpha Energy Calibration of CR-39 Using an Am-241 Alpha Source}

\begin{figure}
  	\centering
    	\includegraphics[width=0.3\columnwidth]{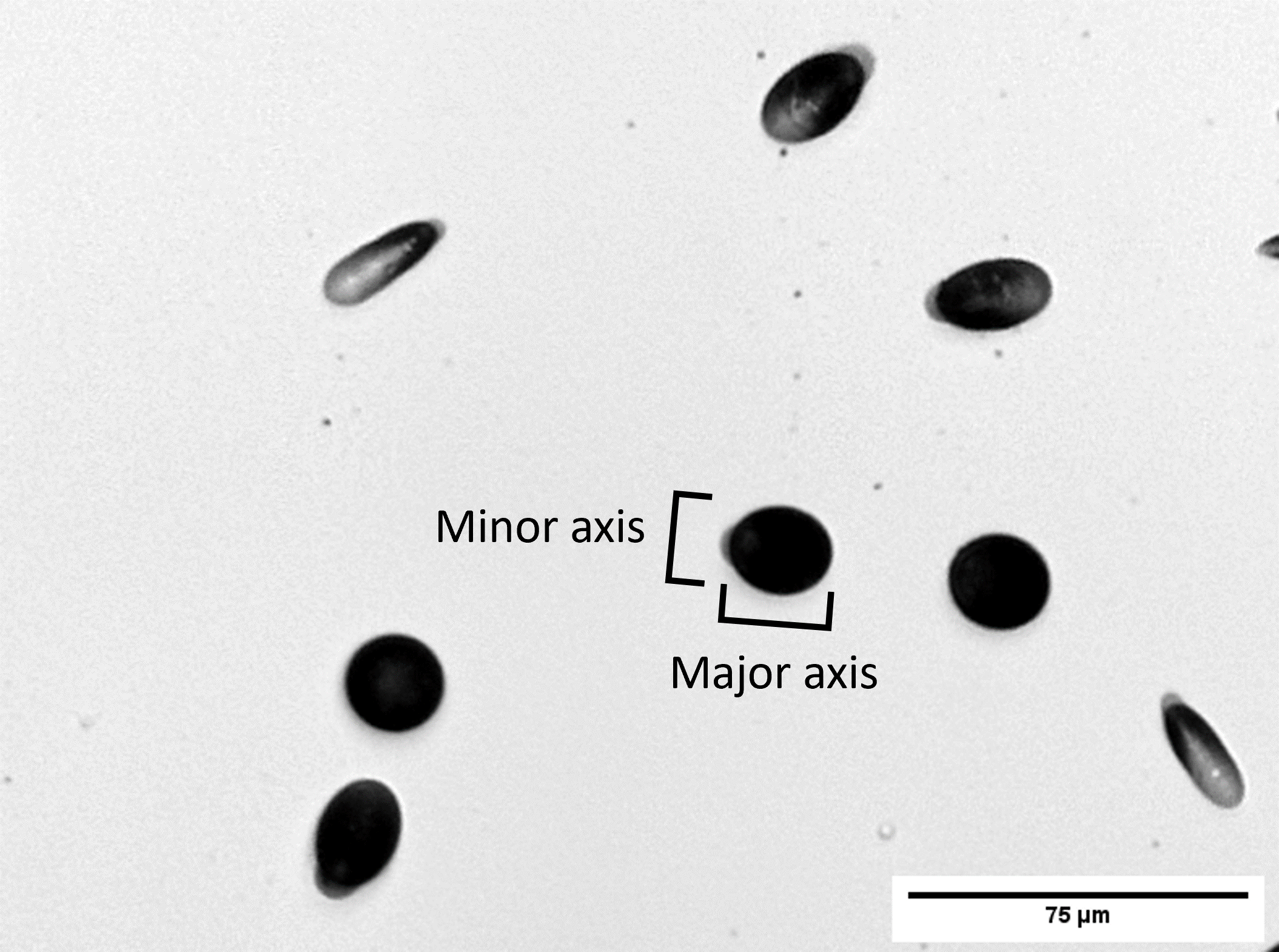}
    	\caption{\label{fig:CR-39 - bare Am-241 tracks}Tracks due to alphas emitted from the \ce{^{241}Am} source placed directly on the CR-39.}
\end{figure}

A 0.12 \textmu Curie \ce{^{241}Am} unsealed source from Isotope Products Laboratories was used to calibrate the CR-39 response to alpha particles with an etching procedure of 1 hr in 6.25 M NaOH at $98 \pm 1 \degree$C. The resulting trends and analytical fits of the empirical data are used to compare the tracks from the plasma experiments. The calibration entails minor axis versus energy, minor axis versus energy for progressive etchings, and minor axis versus etching time, while the next section covers the depth of tracks versus energy. \cref{fig:CR-39 - bare Am-241 tracks} shows tracks due to alphas from the \ce{^{241}Am} source placed directly on the CR-39. The figure is annotated with the minor and major axis of a track. Particles of normal incidence are typically used for calibration, however, often this is not available as depicted by the elongated tracks due to particles incident with high angles in \cref{fig:CR-39 - bare Am-241 tracks}. Because of this dependence of track diameter on incident angle, the track minor axis was chosen as the calibration parameter since it depends less on the angle than the major axis \cite{Garcia2016}. A circularity parameter of 0.91 was chosen to discriminate from incident angles too far off normal, where circularity is defined as $Circularity = 4 \pi (area/perimeter^2)$. An air gap between the \ce{^{241}Am} source and CR-39 surface was chosen for energy attenuation of 5.486 MeV alphas. Using TRIM, the resulting transmitted energies of 100,000 alphas was calculated over a range of gaps and the data is shown in Table~\ref{Table: air gaps and transmitted energies}. 

\begin{table}
\centering
\caption{\label{Table: air gaps and transmitted energies}Air gaps and corresponding transmitted energies calculated using TRIM.}
\resizebox{0.4\columnwidth}{!}{%
\begin{tabular}{|c|c|c|c|}
\hline
\textbf{Air gap (mm)} & \textbf{Error (mm)} & \textbf{\begin{tabular}[c]{@{}c@{}}Transmitted \\ energy (MeV)\end{tabular}} & \textbf{\begin{tabular}[c]{@{}c@{}}Standard \\ Deviation (MeV)\end{tabular}} \\ \hline
0 & 0.005 & 5.49 & 0.05 \\ \hline
10.54 & 0.005 & 4.52 & 0.02 \\ \hline
20.3 & 0.005 & 3.47 & 0.03 \\ \hline
29.74 & 0.005 & 2.25 & 0.05 \\ \hline
35.03 & 0.005 & 1.27 & 0.07 \\ \hline
37.5 & 0.005 & 0.71 & 0.08 \\ \hline
38.31 & 0.005 & 0.52 & 0.08 \\ \hline
38.96 & 0.005 & 0.37 & 0.07 \\ \hline
39.39 & 0.005 & 0.28 & 0.06 \\ \hline
39.9 & 0.005 & 0.19 & 0.05 \\ \hline
\end{tabular}%
}
\end{table}

\begin{figure}
  	\centering 
    	\includegraphics[width=0.35\columnwidth]{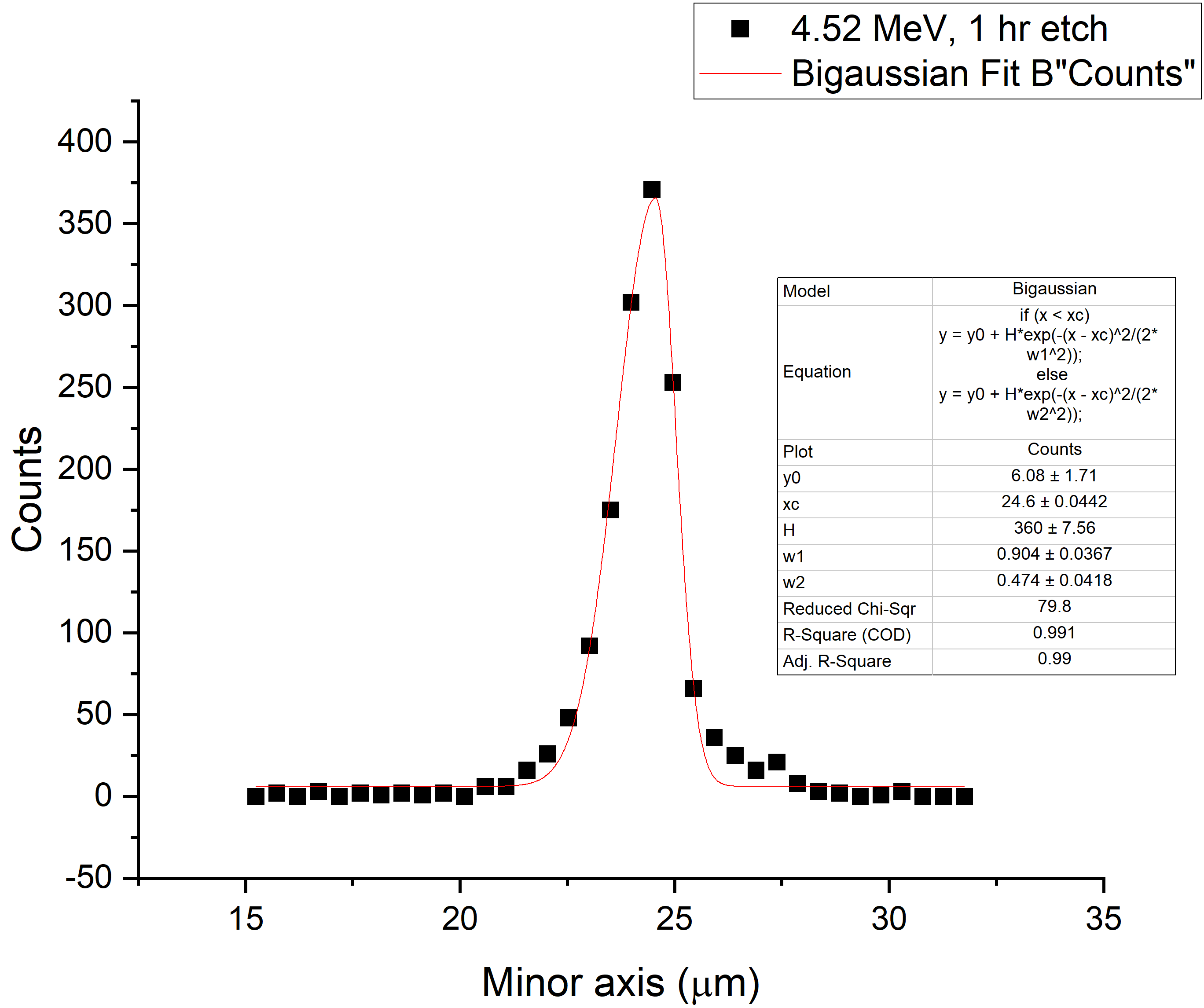}
    	\caption{\label{fig:CR-39 - 10.54mm air gap minor axis peak fit}Bigaussian peak fit of minor axes for tracks corresponding to $4.52 \pm 0.02$ MeV alphas.}
\end{figure}

\begin{figure}
  \centering
      \begin{subfigure}{0.2\textwidth}
        \includegraphics[width=1\textwidth]{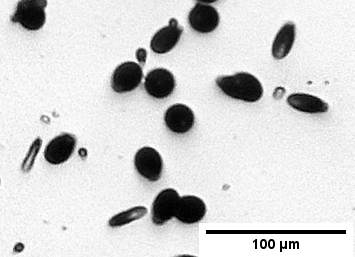}
        \caption{}
      \end{subfigure}
        \begin{subfigure}{0.2\textwidth}
        \includegraphics[width=1\textwidth]{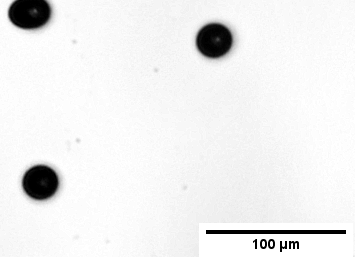}
        \caption{}
      \end{subfigure}
       \begin{subfigure}{0.2\textwidth}
        \includegraphics[width=1\textwidth]{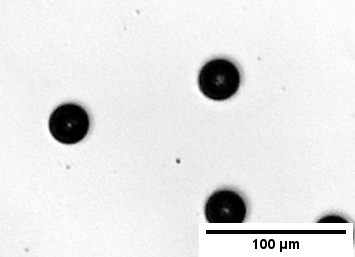}
        \caption{}
      \end{subfigure}   
		\begin{subfigure}{0.2\textwidth}
        \includegraphics[width=1\textwidth]{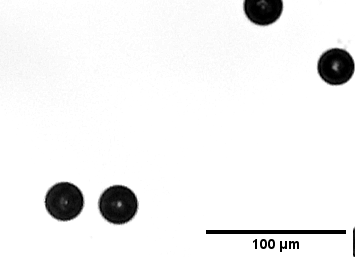}
        \caption{}
      \end{subfigure}        
       \begin{subfigure}{0.2\textwidth}
        \includegraphics[width=1\textwidth]{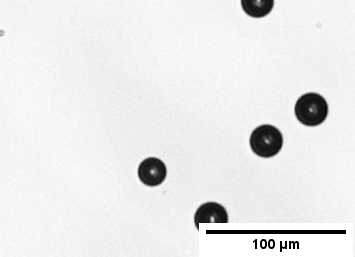}
        \caption{}
      \end{subfigure}      
       \begin{subfigure}{0.2\textwidth}
        \includegraphics[width=1\textwidth]{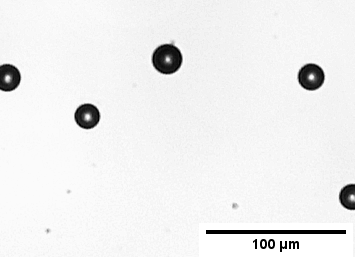}
        \caption{}
      \end{subfigure}      
		\begin{subfigure}{0.2\textwidth}
        \includegraphics[width=1\textwidth]{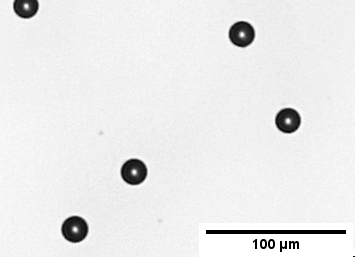}
        \caption{}
        \end{subfigure}  
       \begin{subfigure}{0.2\textwidth}
        \includegraphics[width=1\textwidth]{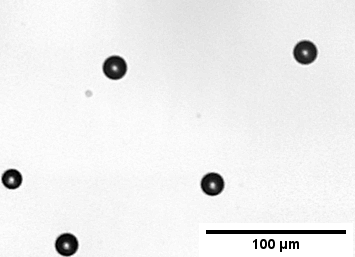}
        \caption{}
      \end{subfigure}      
       \begin{subfigure}{0.2\textwidth}
        \includegraphics[width=1\textwidth]{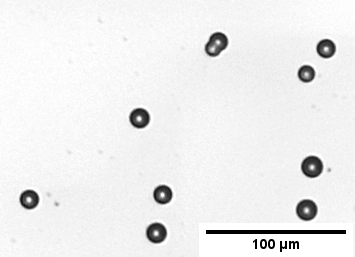}
        \caption{}
      \end{subfigure}      
       \begin{subfigure}{0.2\textwidth}
        \includegraphics[width=1\textwidth]{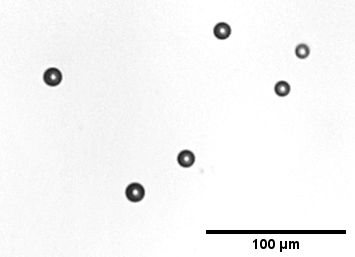}
        \caption{}
      \end{subfigure}      
      \caption{Tracks due to alphas with energy (a) 5.49 MeV, (b) 4.52 MeV, (c) 3.47 MeV, (d) 2.25 MeV, (e) 1.27 MeV, (f) 0.71 MeV, (g) 0.52 MeV, (h) 0.37 MeV, (i) 0.28 MeV, and (j) 0.19 MeV.}
      \label{fig:CR-39 - example images of Am-241 tracks}
\end{figure}

After irradiation, the samples were etched, imaged, and analyzed. An example of the track minor axes histogram fit with a biagaussian curve is shown in \cref{fig:CR-39 - 10.54mm air gap minor axis peak fit}. The bigaussian curve fit was used instead of a gaussian curve to counteract deviations in the distance between the source and the CR-39 surface and to account for the \ce{^{241}Am} source's polyenergetic decay scheme, i.e., 5.486 MeV (85.2$\%$) and 5.443 MeV (12.8$\%$). Examples of tracks for each energy is shown in \cref{fig:CR-39 - example images of Am-241 tracks}. The change in contrast is noticeable as the energy decreases and the tracks become shallower.

\begin{figure}
  	\centering
    	\includegraphics[width=0.4\columnwidth]{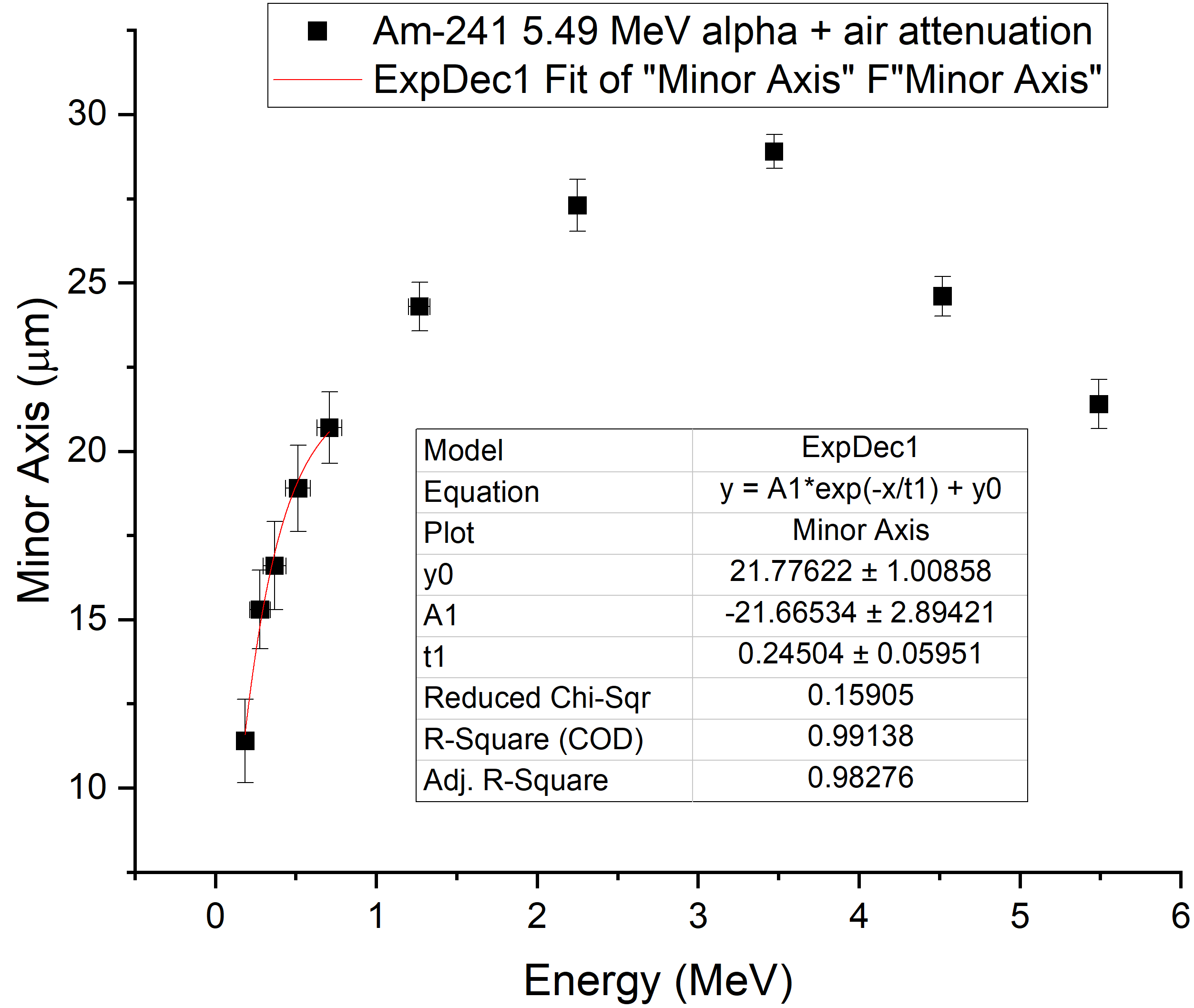}
	\caption{\label{fig:CR-39 - Minor axis vs energy alphas}Minor axis versus energy for \ce{^{241}Am} alphas attenuated with air gaps of varying distance. Exponential decay fit for the lowest five energy values.}
\end{figure}

The minor axis data is then used to make the final minor axis versus incident energy for alpha particles shown in \cref{fig:CR-39 - Minor axis vs energy alphas}. This data is analogous to what was done by Duan et al. in \cite{xiaojiaoCalibrationCR39Monoenergetic2009} for monoenergetic protons of (a) 20 keV, (b) 40 keV, (c) 60 keV, (d) 80 keV, (e) 100 keV, (f) 220 keV, (g) 320 keV, (h) 420 keV, (i) 520 keV, (j) 620 keV, (k) 820 keV, and (l) 1200 keV then etched in 6 M NaOH at $70 \degree$C for 9 hours. It was noted that the optical limit for observing tracks from protons for their etching procedure and optical imaging was $\sim$20 keV. The optical limit of alpha tracks for the imaging system used in this work with a resolving power of 2.58 \textmu m was then 31 keV. 

\subsection*{\label{Sec: Am-241 progressive etching}Progressive Etching of the Five Lowest Energy Tracks}

\begin{figure}
  \centering
      \begin{subfigure}{0.2\textwidth}
        \includegraphics[width=1\textwidth]{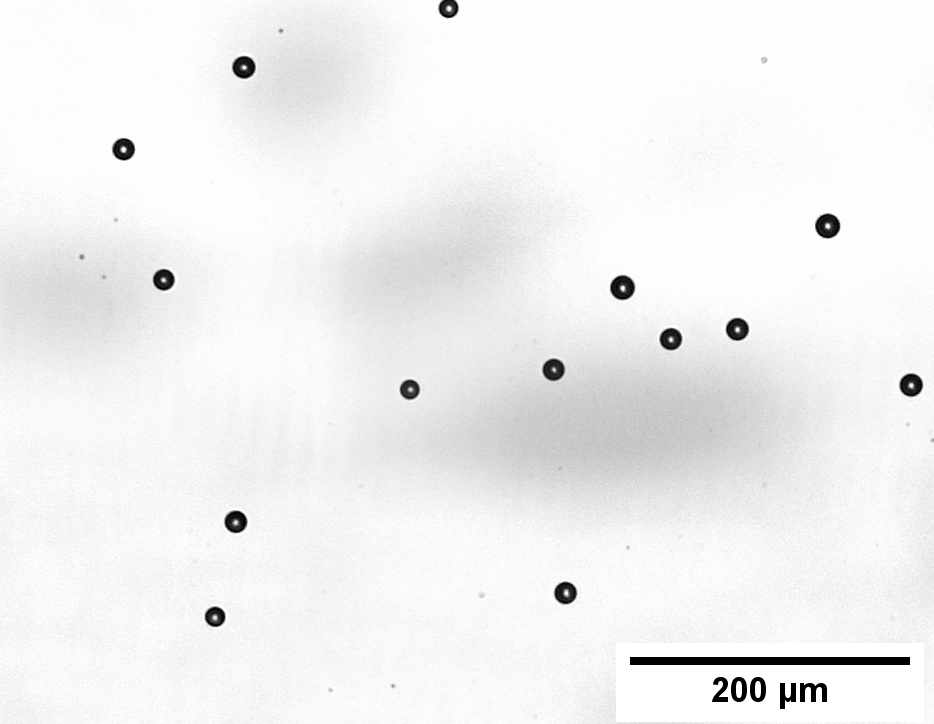}
        \caption{}
      \end{subfigure}
      \begin{subfigure}{0.2\textwidth}
        \includegraphics[width=1\textwidth]{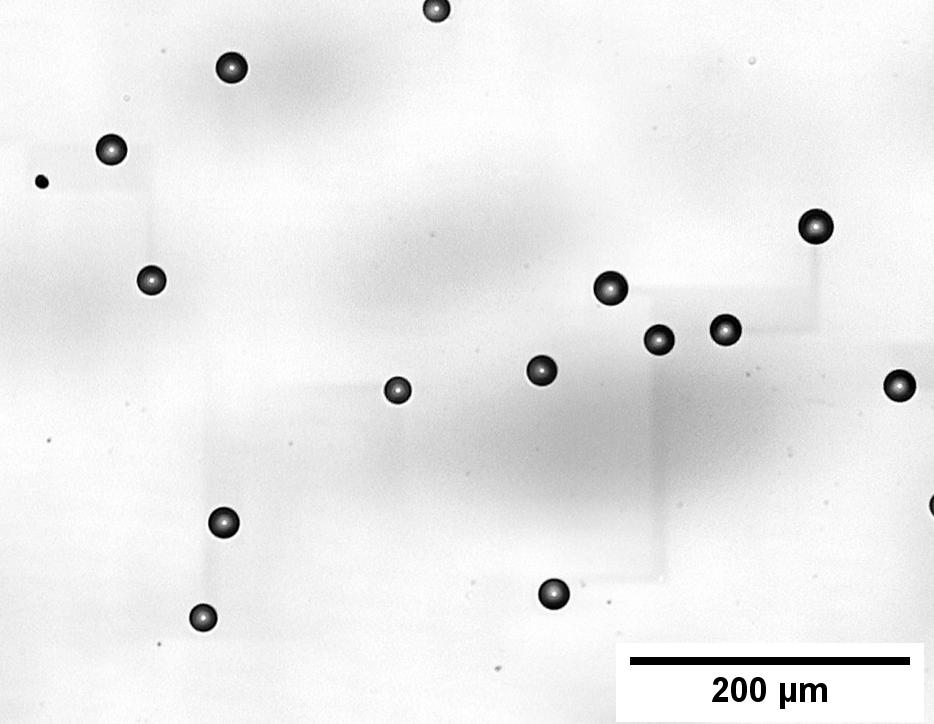}
        \caption{}
      \end{subfigure}
       \begin{subfigure}{0.2\textwidth}
        \includegraphics[width=1\textwidth]{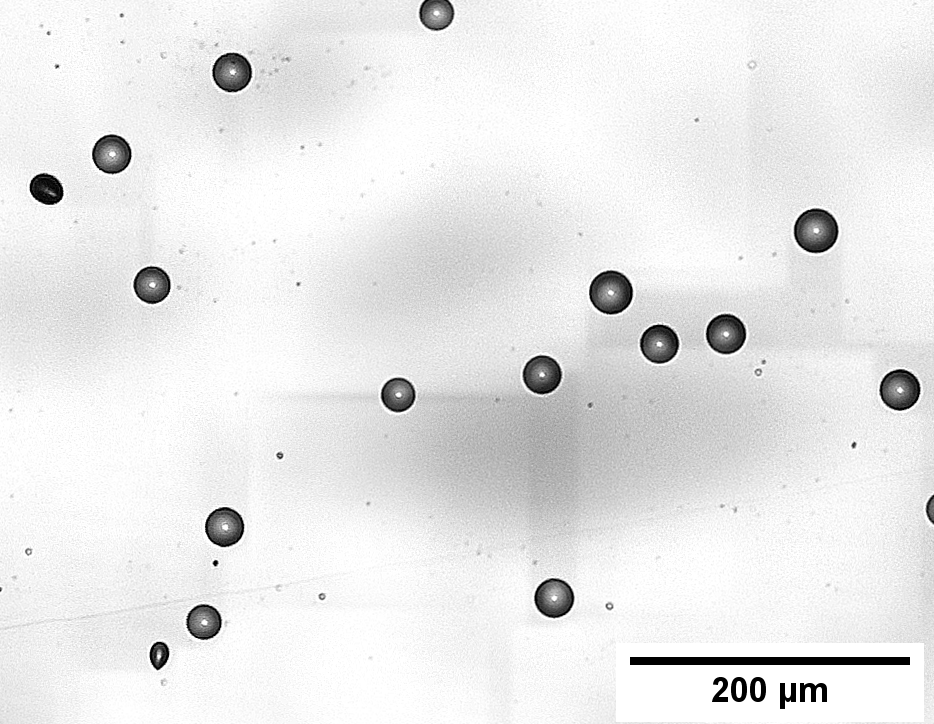}
        \caption{}
      \end{subfigure}   
		\begin{subfigure}{0.2\textwidth}
        \includegraphics[width=1\textwidth]{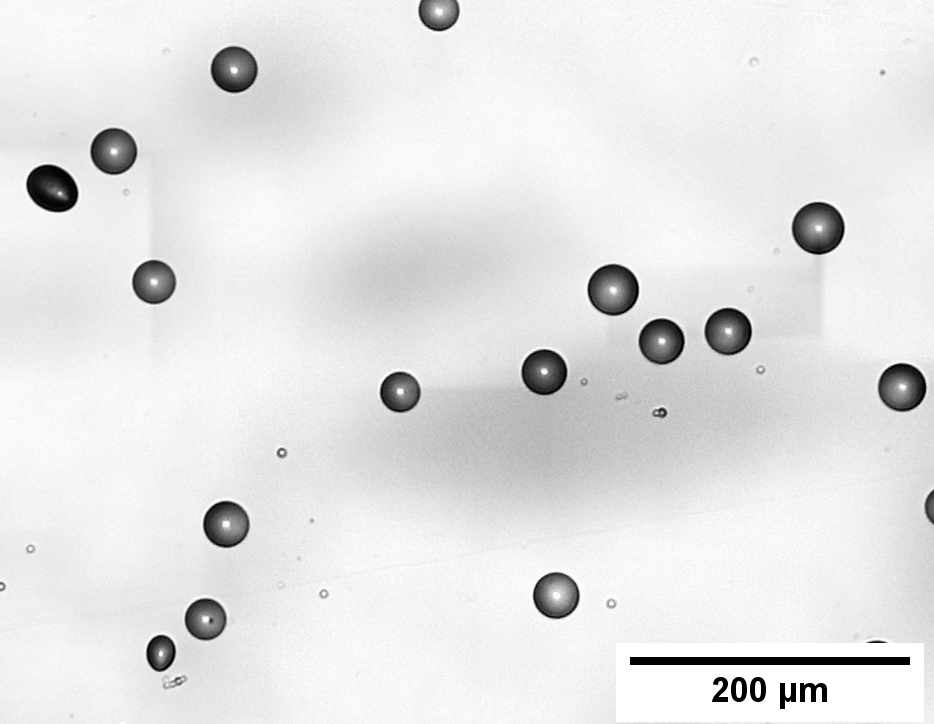}
        \caption{}
      \end{subfigure}        
       \begin{subfigure}{0.2\textwidth}
        \includegraphics[width=1\textwidth]{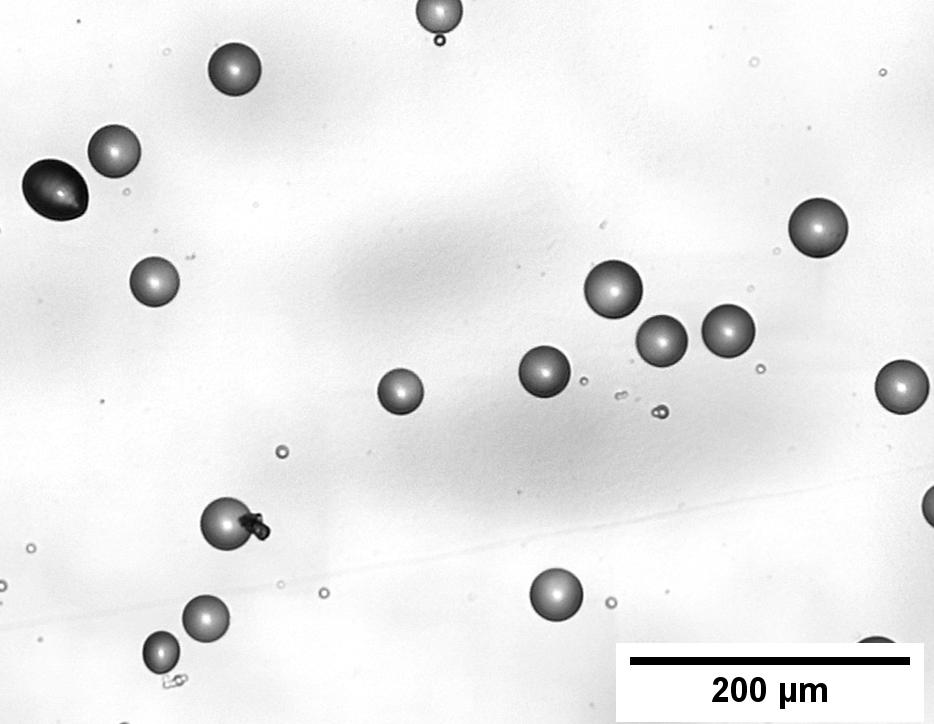}
        \caption{}
      \end{subfigure}                    
      \caption{Tracks due to 0.37 MeV alphas after etching in 6.25 M NaOH at $98\pm1\degree$C for (a) 1 hr, (b) 1 hr + 1 hr, (c) 1 hr + 1 hr + 1 hr, (d) 1 hr + 1 hr + 1 hr + 1 hr, and (e) 1 hr + 1 hr + 1 hr + 1 hr + 1 hr.}
      \label{fig:CR-39 - 0.37 MeV tracks progressive etch}
\end{figure}

\begin{figure}
  \centering
      \begin{subfigure}{0.32\textwidth}
        \includegraphics[width=1\textwidth]{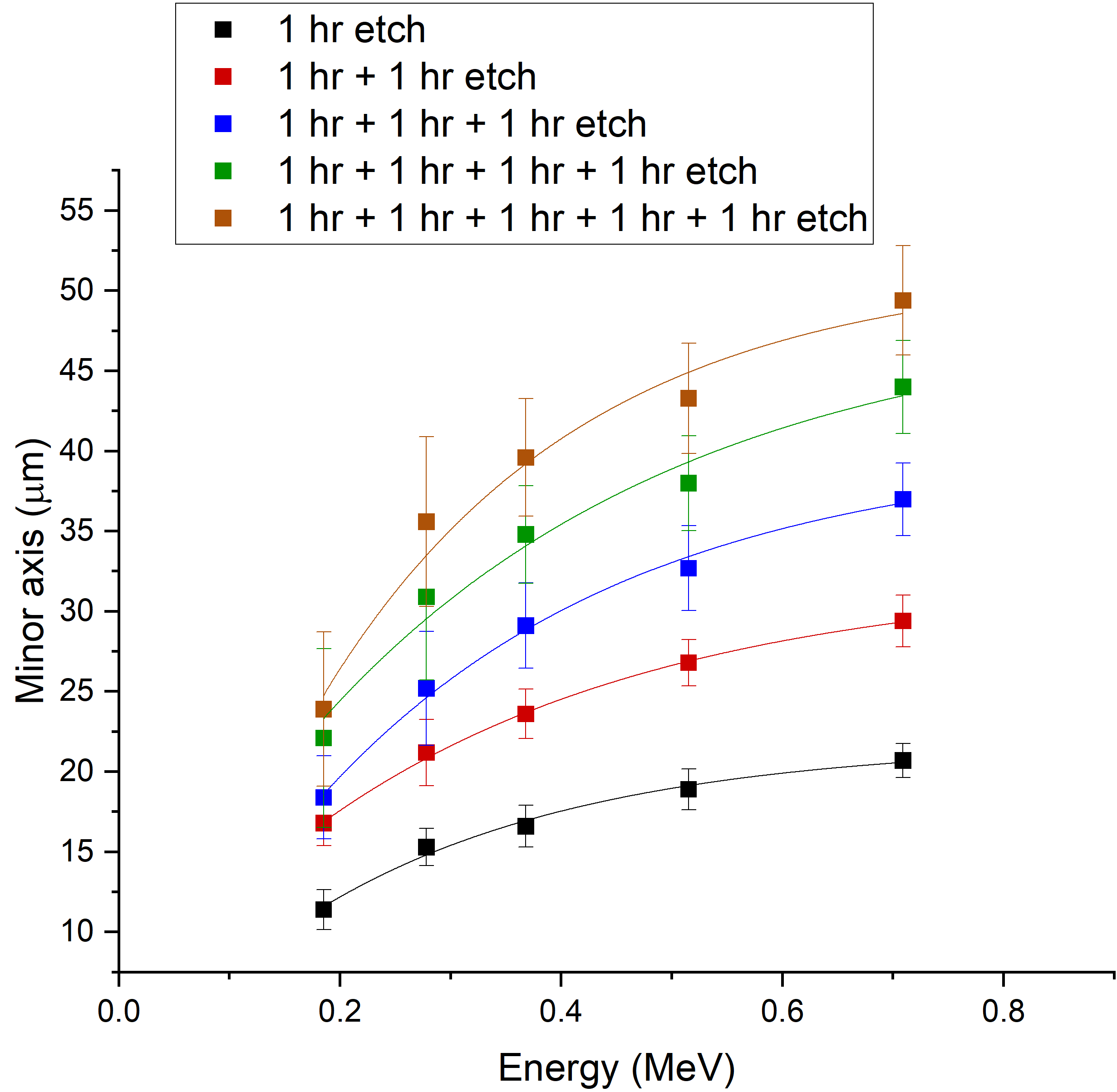}
        \caption{}
      \end{subfigure}
      \begin{subfigure}{0.32\textwidth}
        \includegraphics[width=1\textwidth]{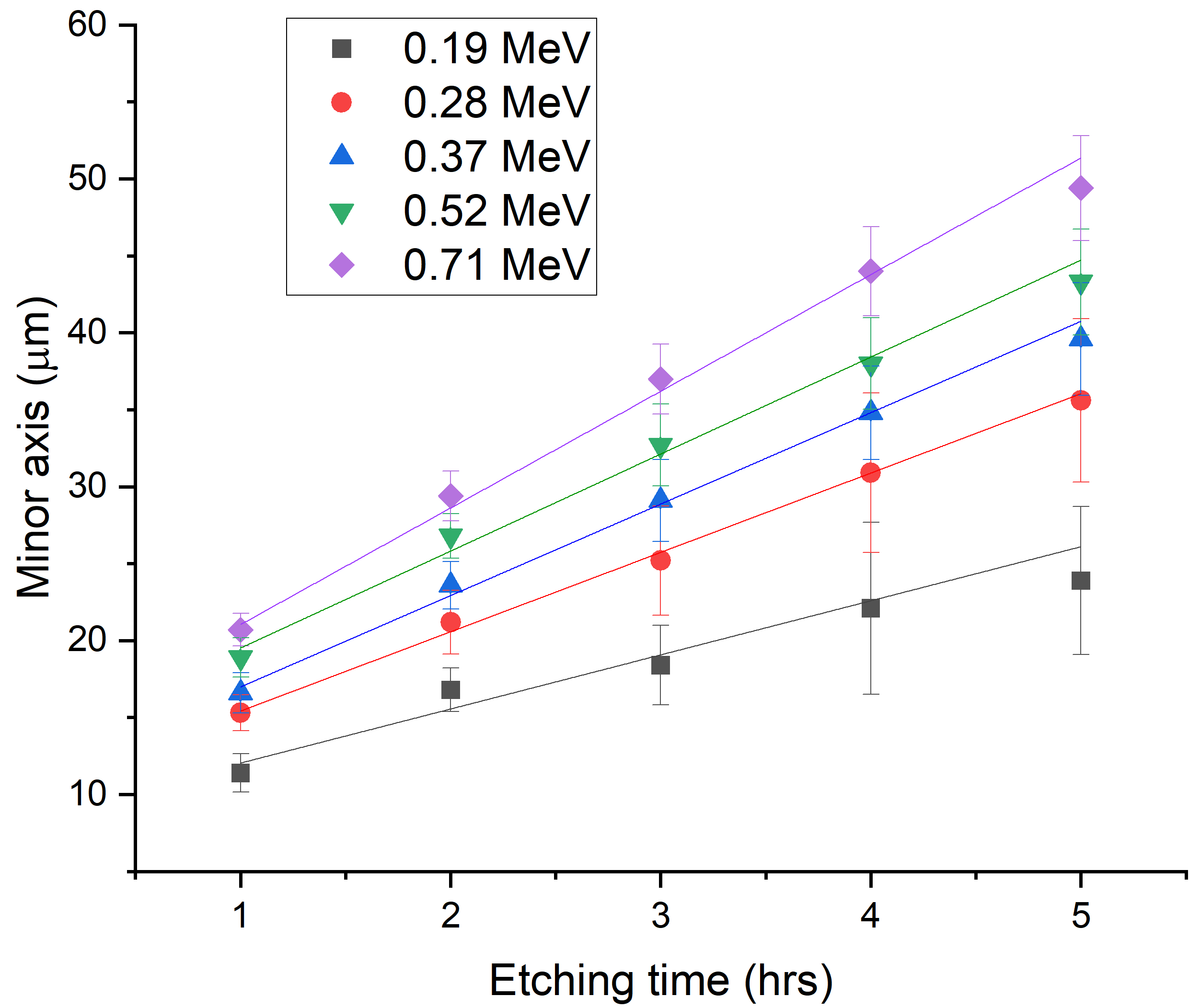}
        \caption{}
      \end{subfigure}
       \begin{subfigure}{0.32\textwidth}
        \includegraphics[width=1\textwidth]{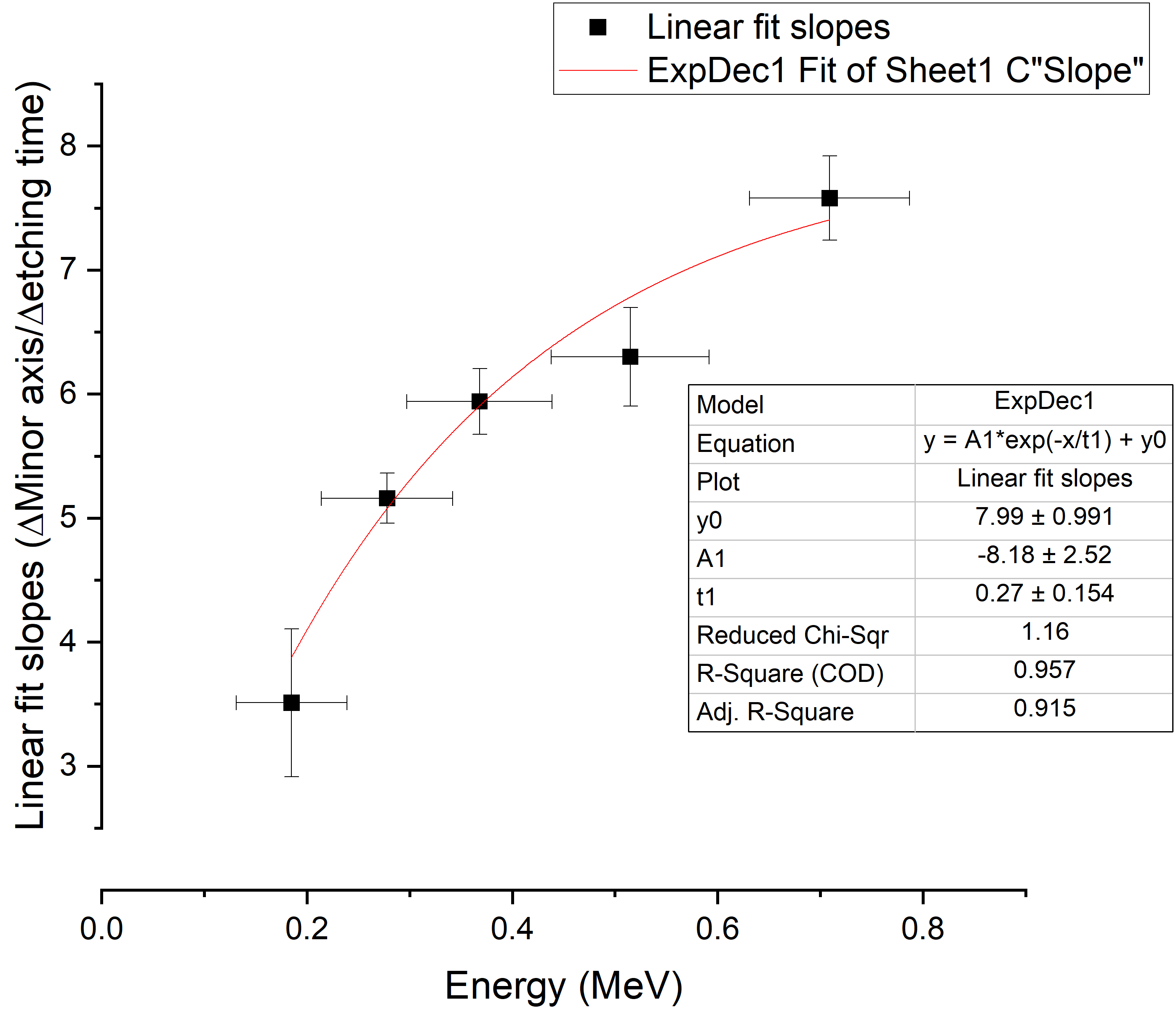}
        \caption{}
      \end{subfigure}   
      \caption{(a) Minor axis versus energy for progressive etches. The error bars for the energy were removed for ease of viewing. (b) Minor axis versus etching time for all energies. (c) Slope of the minor axis versus energy linear fits for all energies.}
      \label{fig:CR-39 - Minor axis vs energy minor axis vs etching time slope of minor axis versus energy}
\end{figure}

As the tracks produced from the discharge experiments had diameters $\sim$8 \textmu m and low contrast indicative of a shallow track, the lowest five energy tracks were of focus for progressive etchings. Progressive etching of tracks is another tool in examining track characteristics and to determine the particle type. Examples of the tracks due to 0.37 MeV alphas as they progressed are shown in \cref{fig:CR-39 - 0.37 MeV tracks progressive etch}. The minor axis versus energy was plotted for each etching interval and then fit to an exponential decay - all plotted together in \cref{fig:CR-39 - Minor axis vs energy minor axis vs etching time slope of minor axis versus energy}a. The same exponential decay shape was found for all etching interval data sets. Next, the minor axis versus etching time for each energy was explored, compiled in \cref{fig:CR-39 - Minor axis vs energy minor axis vs etching time slope of minor axis versus energy}b. The minor axis was found to increase linearly with etching time, which means the latent damage in the polymer was still being preferentially etched compared to the bulk. The linear relationship between track minor axis and etching time was also found in \cite{Khan1983}. Last, the linear slope for each energy interval was fit to an exponential decay as another comparison measure - \cref{fig:CR-39 - Minor axis vs energy minor axis vs etching time slope of minor axis versus energy}c. 

\subsection*{\label{Sec: Track depths}3D Laser Scanning Microscope - Track Depths and Minor Axis Diameters}

This section covers the use of a Keyence 3D Laser Scanning Microscope - VK-X1000. It was used to investigate the track depths for the five lowest energies, again 0.71 Mev, 0.52 MeV, 0.37 MeV, 0.28 MeV, and 0.19 MeV. The measurements were done manually which limits the selection of data to roughly 30 tracks for each energy interval. Due to the small sample size, the data was fit to a broad normal distribution instead of the bigaussian peak fit used in the previous automated analysis pipeline.

\begin{figure}
  \centering
       \begin{subfigure}{0.22\textwidth}
        \includegraphics[width=1\textwidth]{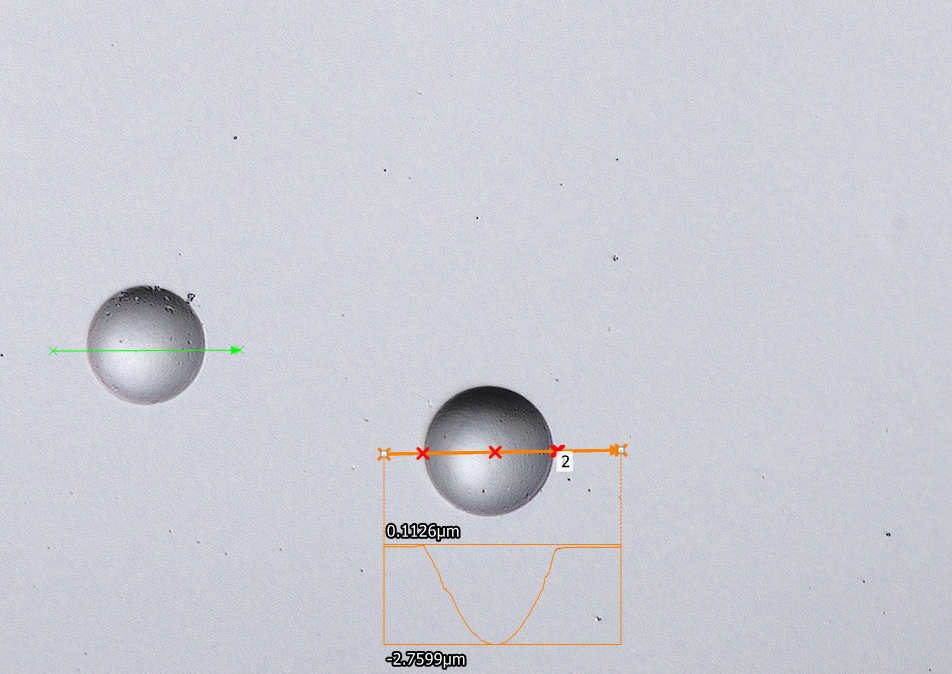}
        \includegraphics[width=1\textwidth]{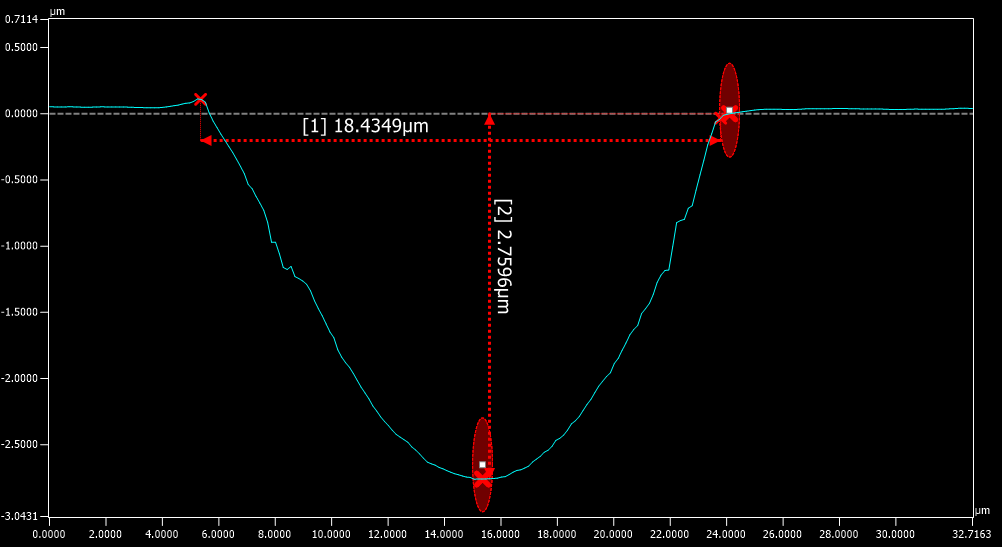}
        \caption{}
      \end{subfigure} 
      \begin{subfigure}{0.22\textwidth}
        \includegraphics[width=1\textwidth]{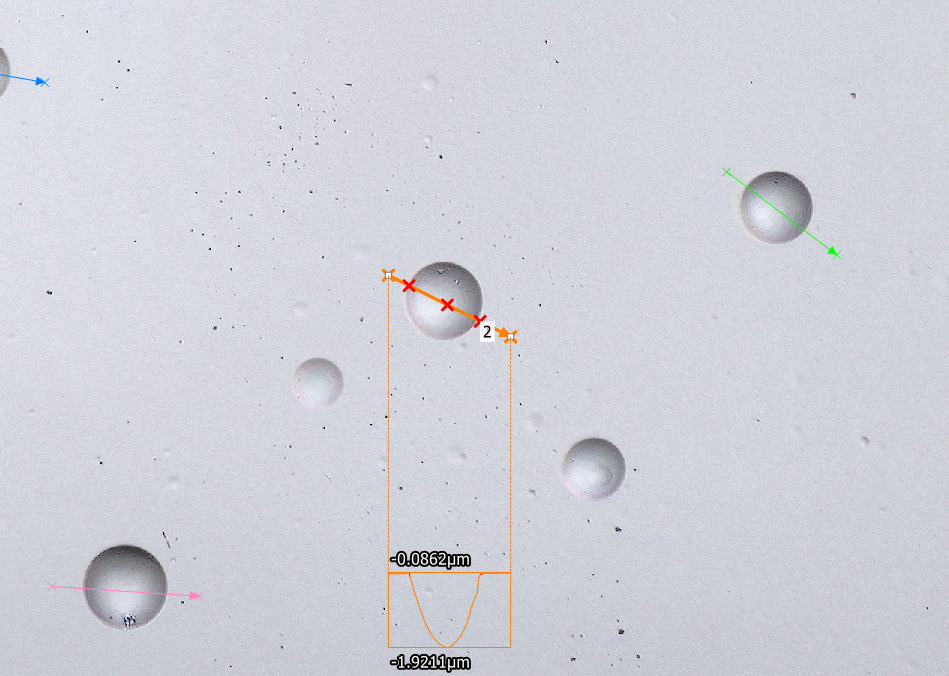}
        \includegraphics[width=1\textwidth]{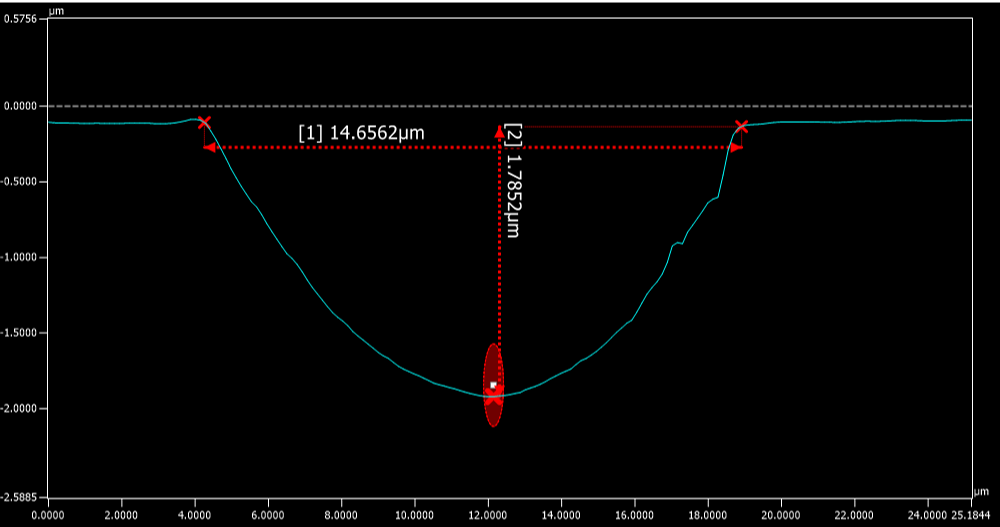}
        \caption{}
        \end{subfigure}         
      \caption{Examples of tracks and their depth profiles taken using the Keyence VK-X1000 for alphas of energy: (a) 0.52 MeV and (b) 0.28 MeV.} 
      \label{fig:CR-39 - example images of Am-241 track depth profiles}
\end{figure}

\begin{figure}
  \centering
      \begin{subfigure}{0.4\textwidth}
        \includegraphics[width=1\textwidth]{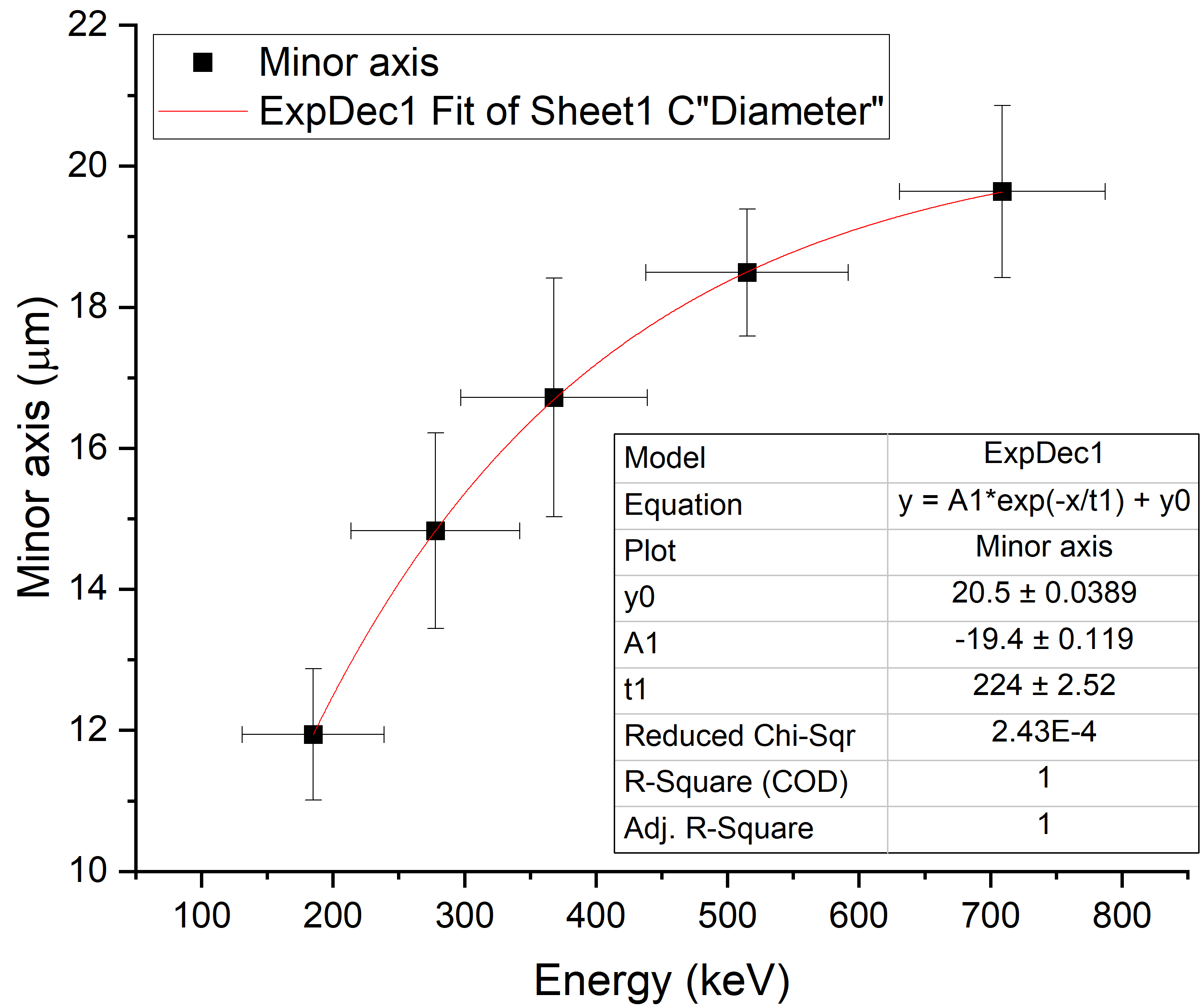}
        \caption{}
      \end{subfigure}
      \begin{subfigure}{0.4\textwidth}
        \includegraphics[width=1\textwidth]{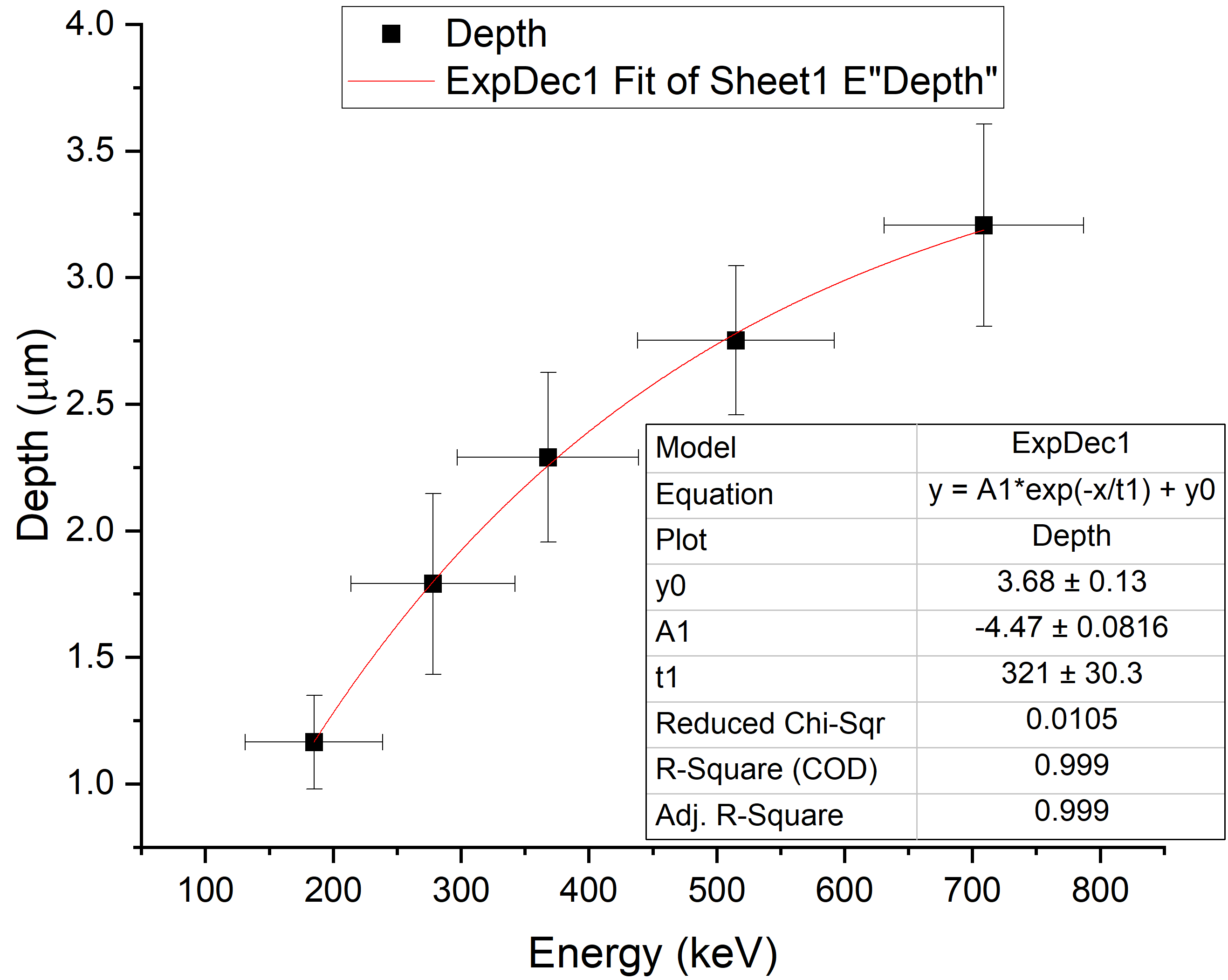}
        \caption{}
      \end{subfigure}            
        \begin{subfigure}{0.4\textwidth}
        \includegraphics[width=1\textwidth]{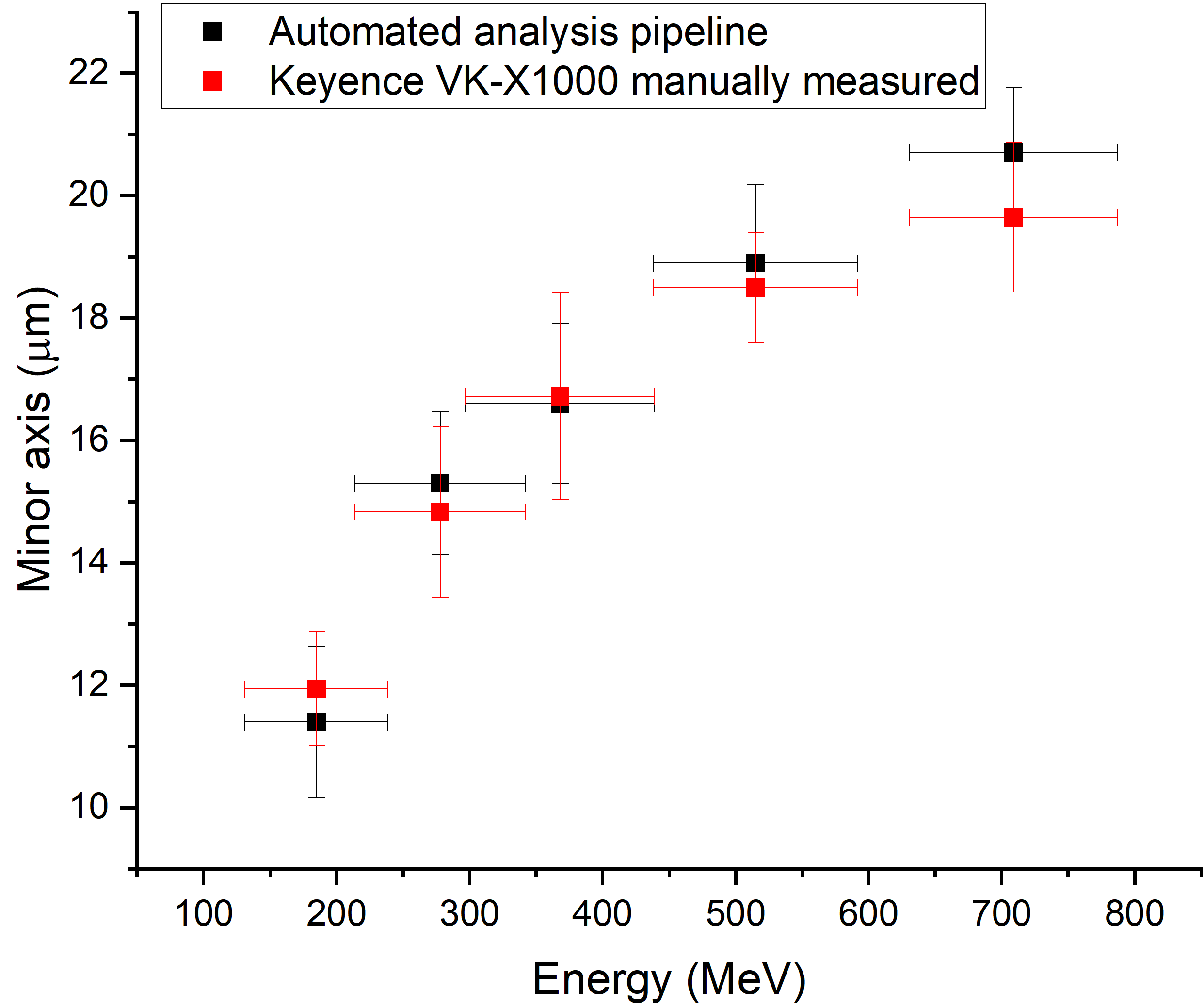}
        \caption{}
      \end{subfigure}  
      \caption{Curve fits for (a) track diameter versus energy and (b) track depth versus energy. (c) Track minor axis versus energy for the automated analysis pipeline and Keyence VK-X1000 data.}
      \label{fig:CR-39 - Keyence Am-241 track depth and diameter vs energy}
\end{figure}

Examples of chosen track depth profiles from 0.52 MeV and 0.28 MeV alphas are shown in \cref{fig:CR-39 - example images of Am-241 track depth profiles}. The track depth versus energy along with measured diameter versus energy are then shown in \cref{fig:CR-39 - Keyence Am-241 track depth and diameter vs energy}. In this energy regime, incident angle, and etching procedure, the diameter is synonymous with the minor axis of the track. The diameter versus energy data for the manual Keyence measurements aligns well with the automated analysis pipeline - \cref{fig:CR-39 - Keyence Am-241 track depth and diameter vs energy}c.

\section*{\label{Sec: Experimental Results}Experimental Results}

\subsection*{\label{Subsec: Hydrogen and helium control tests}Comparison of hydrogen and helium controls and deuterium track densities}

\begin{figure}[H]
\centering
\begin{subfigure}{0.3\textwidth}
\includegraphics[width=1\textwidth]{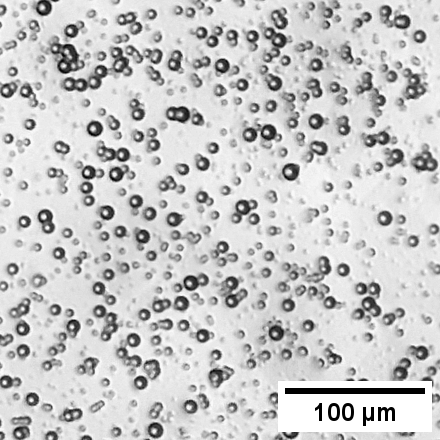}
\caption{}
\end{subfigure}
\begin{subfigure}{0.3\textwidth}
\includegraphics[width=1\textwidth]{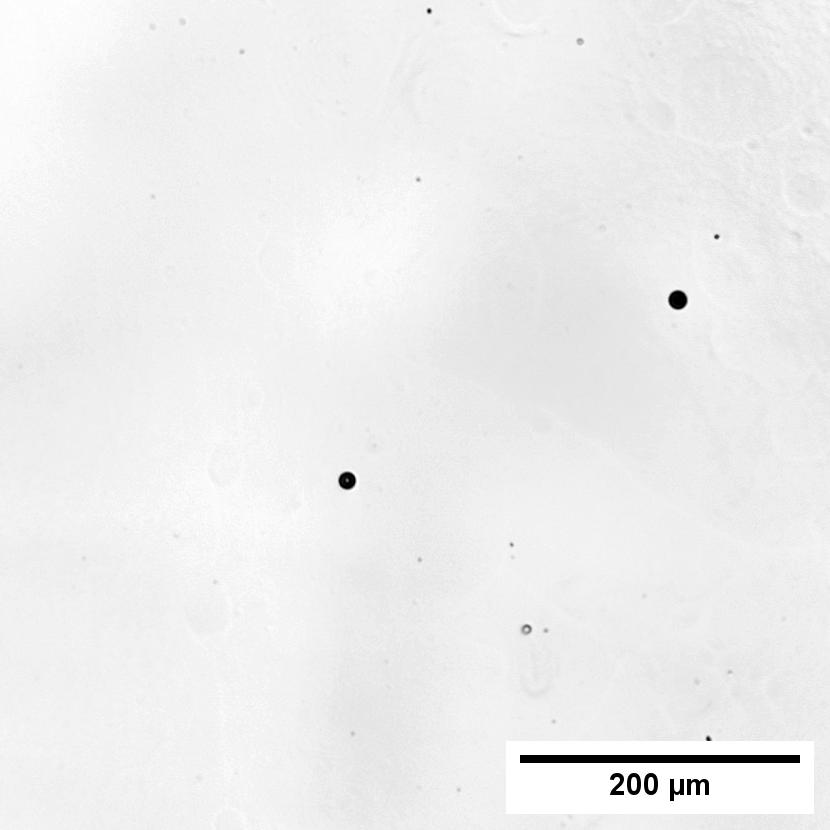}
\caption{}
\end{subfigure}
\begin{subfigure}{0.3\textwidth}
\includegraphics[width=1\textwidth]{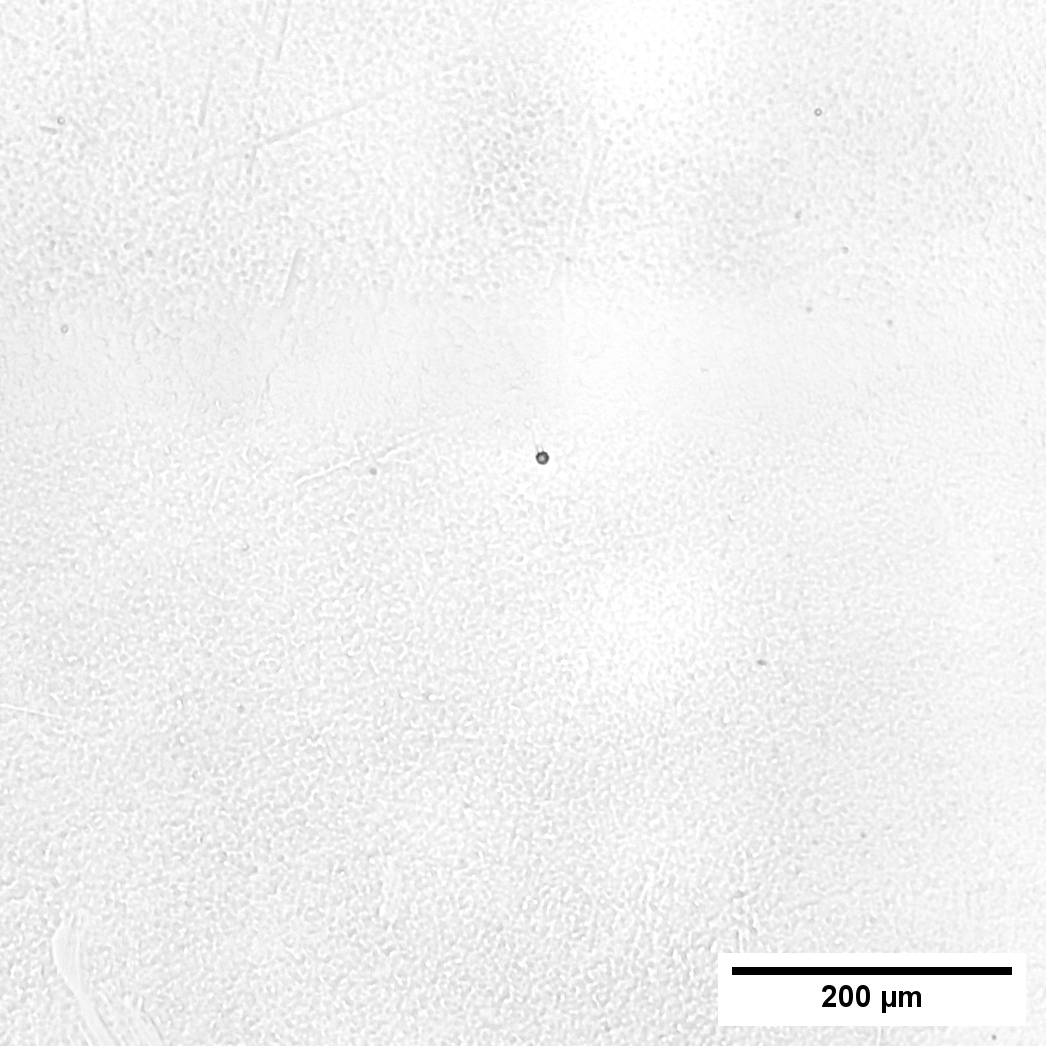}
\caption{}
\end{subfigure}
\begin{subfigure}{0.3\textwidth}
\includegraphics[width=1\textwidth]{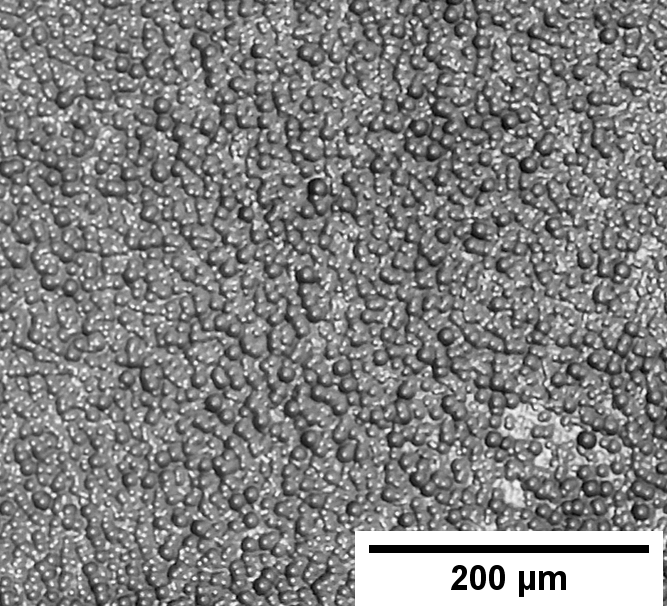}
\caption{}
\end{subfigure}
\caption{Example of tracks after discharges with palladium electrodes. (a) and (d) deuterium, (b) hydrogen, and (c) helium.}
\label{fig:CR-39 - example of plasma tracks}
\end{figure}

\cref{fig:CR-39 - example of plasma tracks} shows an example of tracks after discharges with deuterium, hydrogen, and helium. The parameters for (a), (b), and (c) were 10 Torr, 5 mm electrode gap distance, -500 $\pm$ 100 V, 40 $\pm$ 0.3 mA/cm$^2$, 5 $\times$ 10$^{21}$ ions/cm$^2$ cathode fluence, and z = 5 cm between the cathode and CR-39. \cref{fig:CR-39 - example of plasma tracks}d went to a fluence of $5 \times 10^{22}$ ions/cm$^2$ and shows how the detector eventually saturates with tracks. 

\begin{figure}
  \centering
        \includegraphics[width=0.4\textwidth]{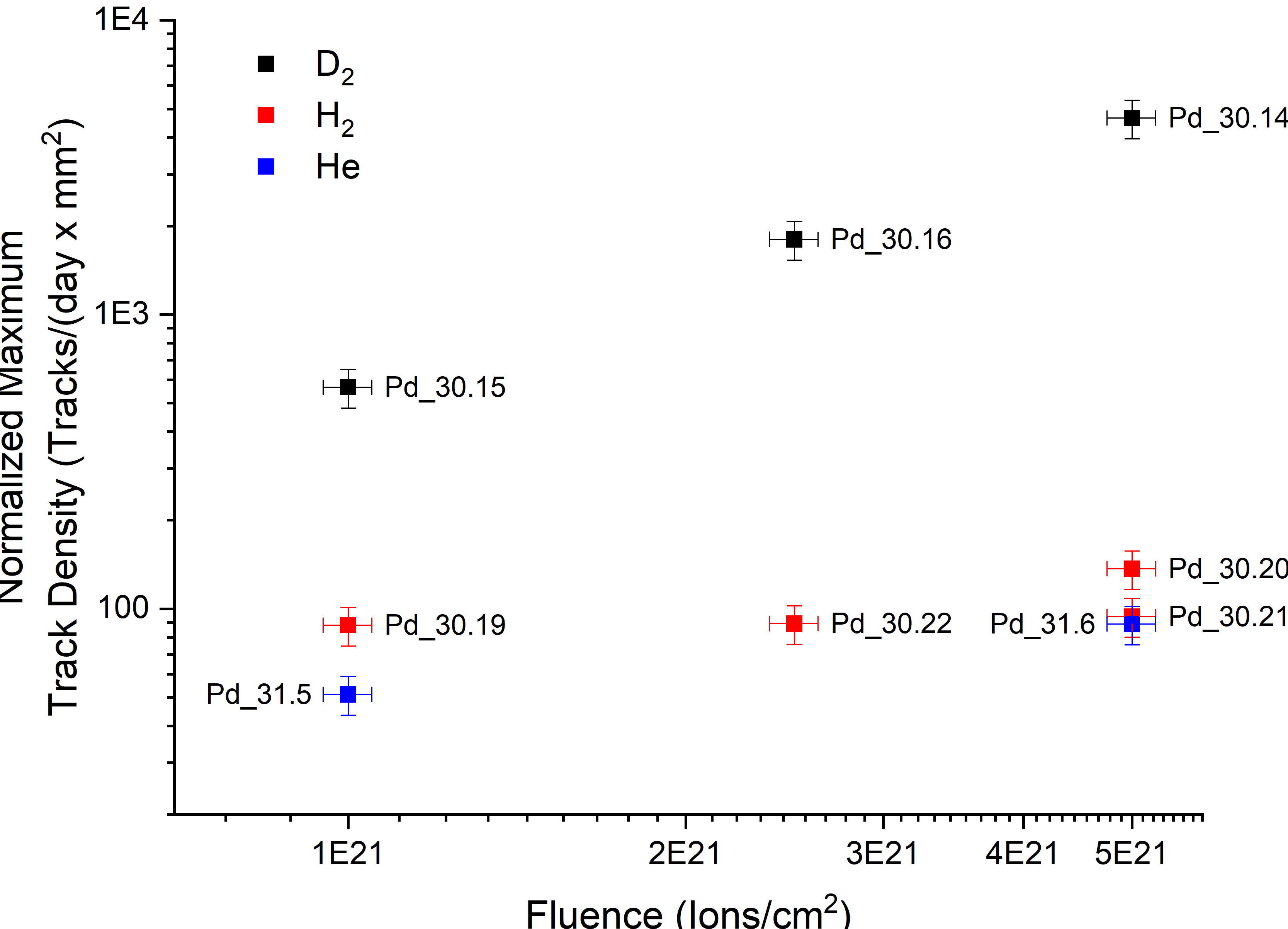}
      \caption{Normalized maximum track density versus fluence for deuterium, hydrogen, and helium discharges at 10 Torr.}
      \label{fig:CR-39 data - HD norm max tracks vs fluence D2 H2 He}
\end{figure}

As the CR-39 chips often had different active areas due to using attenuation foils over sections of the chip, the surface images were discretized to 0.25 x 0.25 mm$^2$ bins and the number of tracks per bin was used to produce track densities for comparison. The track density data was normalized to the total time of exposure from the deleting etch up to the track etch to combat the effect of track densities increasing as their exposure time increased. For example, if it took three days between the deleting etch and track etch, then all the track density data is normalized by three days. The maximum track density normalized to the total exposure time of the detector was chosen as a criterion for comparison between the controls and deuterium tests. The normalized maximum track density versus cathode ion fluence is shown in \cref{fig:CR-39 data - HD norm max tracks vs fluence D2 H2 He}. The data labels are designated as Pd$\_$X.Y where X is the electrode number of an unused As-Received Pd sample and Y is the number of experiments conducted with the same electrode. For example, Pd$\_$30.14 is the 30th Pd electrode that was used in 14 separate plasma experiments. Note that Pd$\_$30 was tested with deuterium and subsequently with hydrogen, which produced decreased track densities. 

The mean normalized maximum track densities for hydrogen and helium were $102 \pm 23$ tracks/mm$^2$/day and $70 \pm 27$ tracks/mm$^2$/day, respectively. The slight discrepancy could be due to the presence of deuterium in the hydrogen supply. \cref{tab:ten highest normalized max track densities} then shows the ten highest normalized maximum track densities and their corresponding electrodes, current densities, and fluences. The mean normalized maximum track density for these was $(8.41 \pm 2.27) \times 10^{3}$ tracks/mm$^2$/day. The highest track density of $1.01 \times 10^4$ tracks/mm$^2$/day was 129 times above the control tests.

\begin{table}
\centering
\caption{\label{tab:ten highest normalized max track densities}Ten highest normalized maximum track densities and their corresponding electrode, current density, and fluence.}
\resizebox{0.45\columnwidth}{!}{%
\begin{tabular}{|l|c|c|c|}
\hline
\textbf{Electrode} & \textbf{\begin{tabular}[c]{@{}c@{}}Current Density\\ (mA/cm$^2$)\end{tabular}} & \textbf{\begin{tabular}[c]{@{}c@{}}Fluence\\ (Ions/cm$^2$)\end{tabular}} & \textbf{\begin{tabular}[c]{@{}c@{}}Normalized Track Density \\ (Tracks/(day x mm$^2$))\end{tabular}} \\\hline
Pd$\_$30.17 & 10 & $5.0 \times 10^{21}$ & $1.1 \times 10^4$ \\ \hline
Pd$\_$32.6 & 80 & $1.1 \times 10^{22}$ & $1.0 \times 10^4$ \\ \hline
Pd$\_$32.3 & 40 & $5.1 \times 10^{21}$ & $9.4 \times 10^3$ \\ \hline
Pd$\_$32.8 & 80 & $5.0 \times 10^{22}$ & $9.4 \times 10^3$ \\ \hline
Pd$\_$33.2 & 40 & $1.7 \times 10^{22}$ & $9.3 \times 10^3$ \\ \hline
Pd$\_$30.18 & 160 & $3.8 \times 10^{22}$ & $9.1 \times 10^3$ \\ \hline
Pd$\_$33.2 & 40 & $1.7 \times 10^{22}$ & $8.8 \times 10^3$ \\ \hline
Pd$\_$32.4 & 80 & $5.0 \times 10^{21}$ & $8.2 \times 10^3$ \\ \hline
Pd$\_$30.14 & 40 & $5.0 \times 10^{21}$ & $4.7 \times 10^3$ \\ \hline
Pd$\_$32.3 & 40 & $5.1 \times 10^{21}$ & $4.0 \times 10^3$ \\ \hline
\end{tabular}%
}
\end{table}

\subsection*{\label{subsec:track diameters and energy estimates}Plasma Track Diameters and Energy Estimates}

\begin{figure}
    \centering
        \begin{subfigure}{0.22\textwidth}
            \includegraphics[width=1\textwidth]{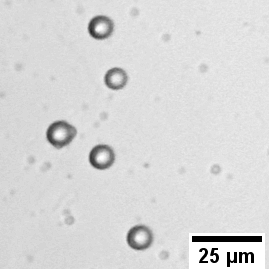}
            \caption{}
        \end{subfigure}
        \begin{subfigure}{0.22\textwidth}
            \includegraphics[width=1\textwidth]{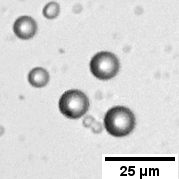}
            \caption{}
        \end{subfigure}
        \begin{subfigure}{0.22\textwidth}
            \includegraphics[width=1\textwidth]{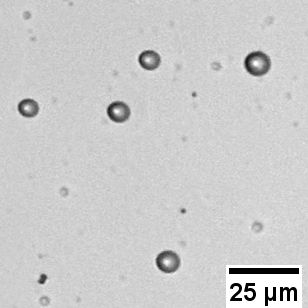}
            \caption{}
        \end{subfigure}   
        \begin{subfigure}{0.22\textwidth}
            \includegraphics[width=1\textwidth]{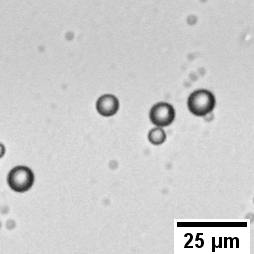}
            \caption{}
        \end{subfigure}       
        \caption{Additional examples of tracks from a deuterium discharge with Pd electrodes.}
        \label{fig:CR-39 - additional examples tracks}
\end{figure}

\begin{figure}
\centering
\includegraphics[width=0.35\columnwidth]{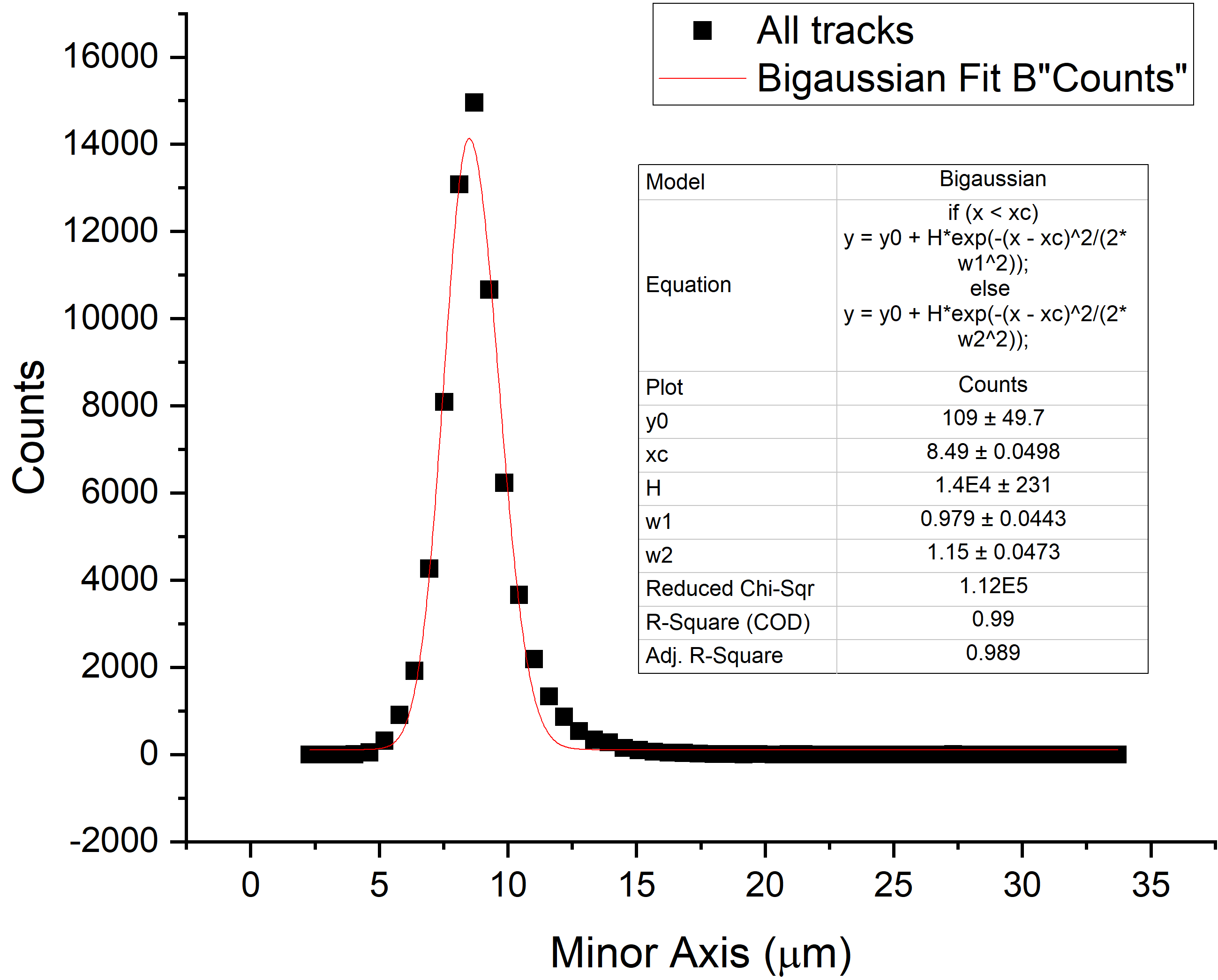}
\caption{\label{fig:CR-39 data - minor axis histogram and peak fit}Bigaussian peak fit of the minor axis histogram data.}
\end{figure}

Tracks produced from a discharge were then compared to the \ce{^{241}Am} calibrations. The discharge parameters were 10 Torr D$_2$, -540 $\pm$ 43 V, 40 $\pm$ 0.2 mA/cm$^2$, $1.74 \times 10^{22}$ ions/cm$^2$ cathode fluence, and z = 5 cm between the cathode and CR-39. Examples of tracks are shown in \cref{fig:CR-39 - additional examples tracks}. The bigaussian peak fit for the track minor axes is shown in \cref{fig:CR-39 data - minor axis histogram and peak fit}. The minor axis peak was $8.49 \pm 0.90$ \textmu m. Using the \ce{^{241}Am} calibration curve, the incident energy was $118 \pm 17$ keV. 

The incident energy is for a particle hitting the CR-39 surface. The particle first has to traverse 5 cm of 10 Torr D$_2$ gas, resulting in energy attenuation. The energy loss was calculated using TRIM with a deuterium gas density of $2.203 \times 10^{-6}$ gram/cm$^3$ and a compound correction of 0.976 from the compound dictionary for mylar\cite{zieglerSRIMStoppingRange2010}. The energy loss then corresponds to the emitted energy required for the particle to have the measured incident energy. The estimated emitted energy was then $139 \pm 19$ keV. 

\subsection*{\label{subsec:track progressive etching}Plasma Track Diameters Progressive Etching}

\begin{figure}
\centering
\begin{subfigure}{0.2\textwidth}
\includegraphics[width=1\textwidth]{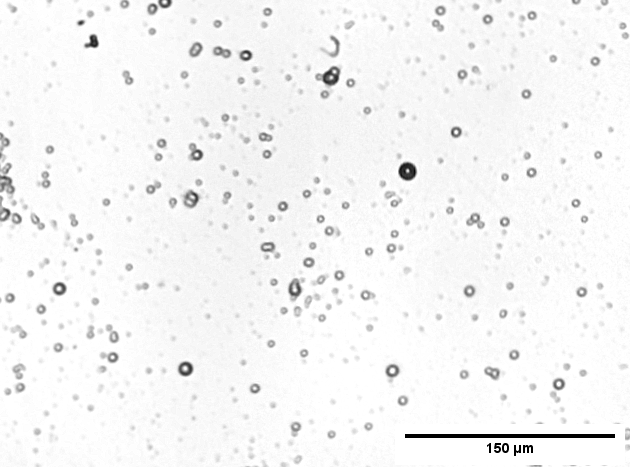}
\caption{}
\end{subfigure}
\begin{subfigure}{0.2\textwidth}
\includegraphics[width=1\textwidth]{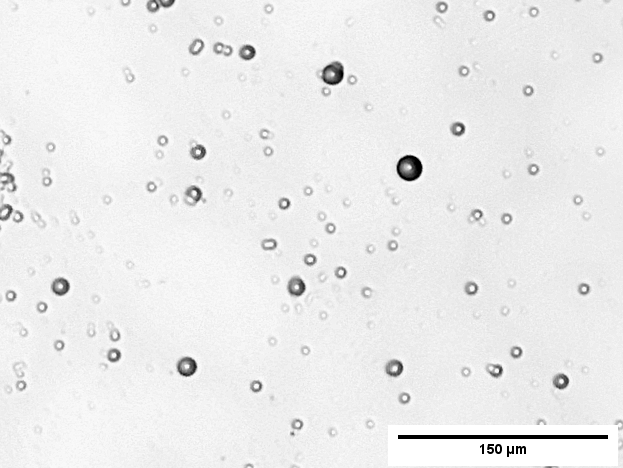}
\caption{}
\end{subfigure}
\begin{subfigure}{0.2\textwidth}
\includegraphics[width=1\textwidth]{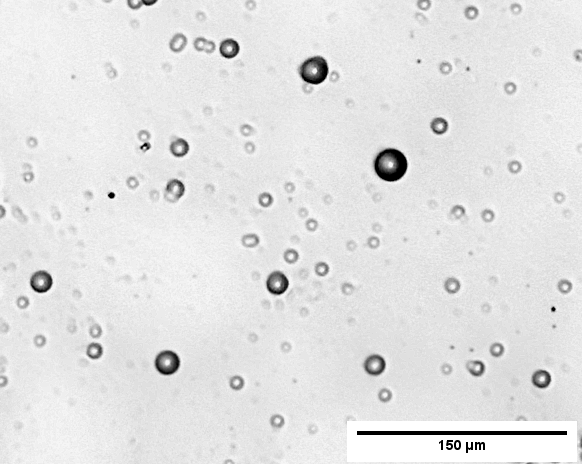}
\caption{}
\end{subfigure}
\begin{subfigure}{0.2\textwidth}
\includegraphics[width=1\textwidth]{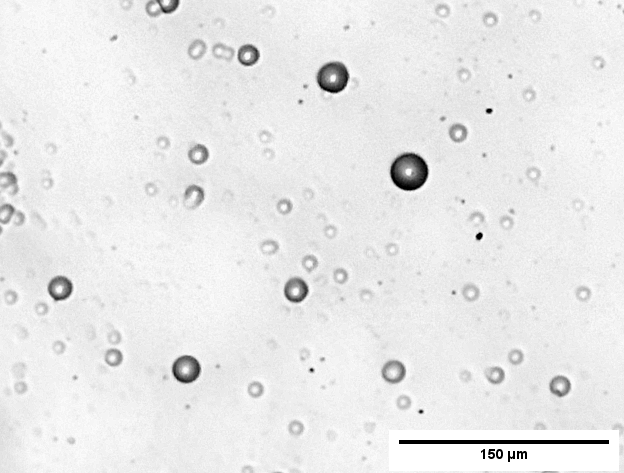}
\caption{}
\end{subfigure}
\begin{subfigure}{0.2\textwidth}
\includegraphics[width=1\textwidth]{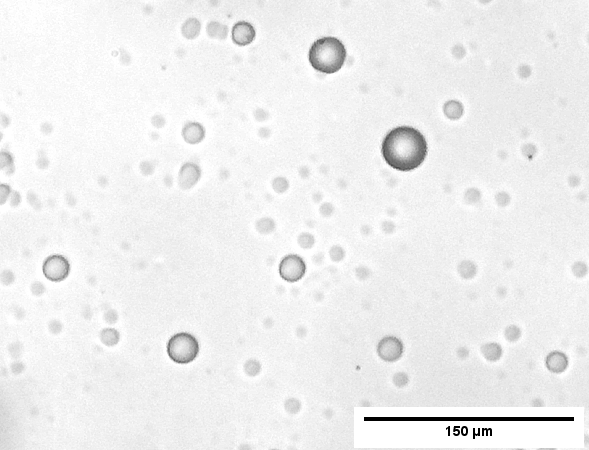}
\caption{}
\end{subfigure}
\caption{Examples of tracks in 00143 for each progressive etch. (a) 1 hr, (b) 1 hr + 1 hr, (c) 1 hr + 1 hr + 1 hr, (d) 1 hr + 1 hr + 1 hr + 1 hr, (e) 1 hr + 1 hr + 1 hr + 1 hr + 1 hr.}
\label{fig:CR-39 00143 progressive etch tracks}
\end{figure}

\begin{figure}
\centering
\includegraphics[width=0.4\columnwidth]{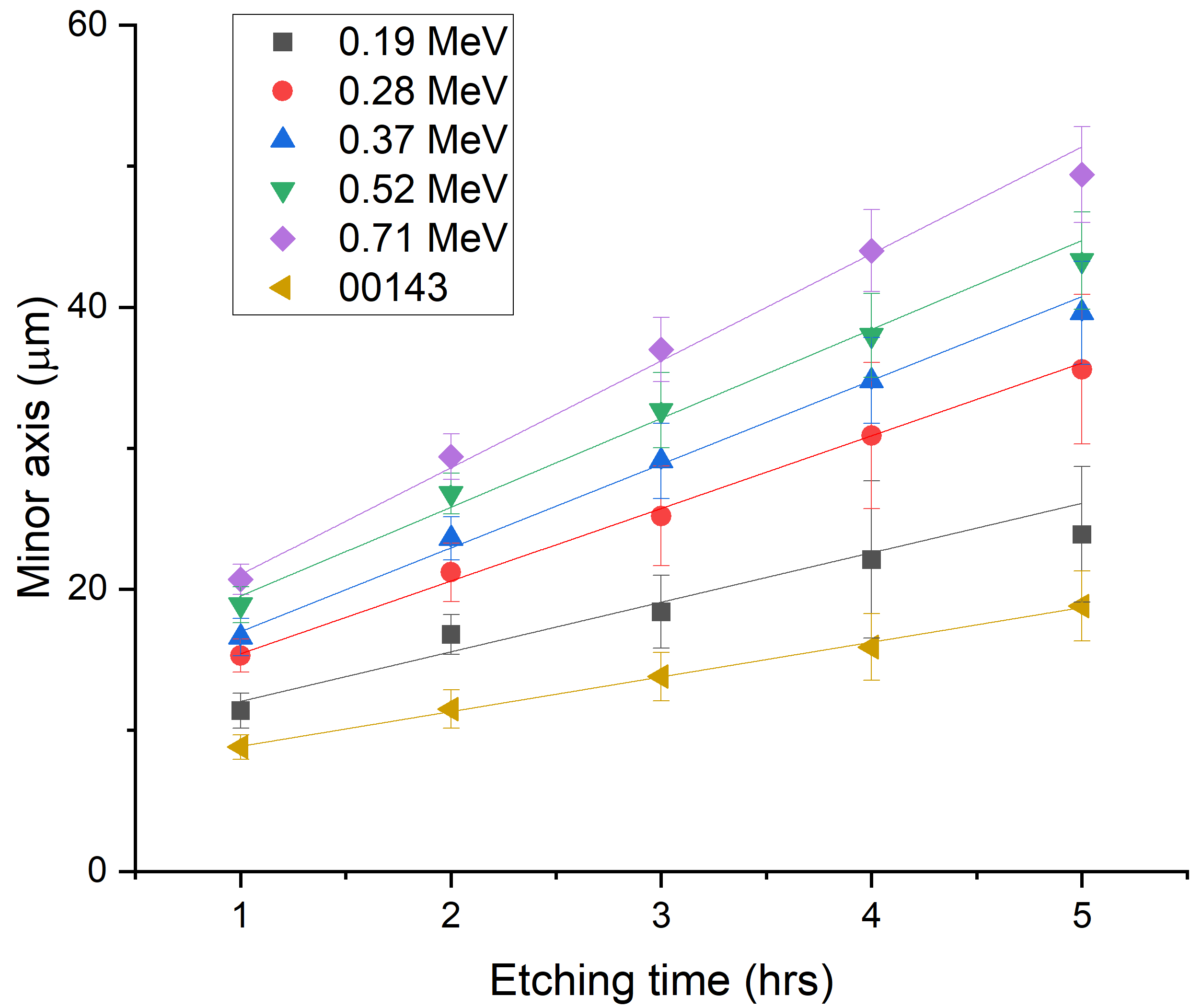}
\caption{\label{fig:CR-39 00143 progressive etch minor axis vs etching time}Progressive etch minor axis versus etching time with the \ce{^{241}Am} standards.}
\end{figure}

Following a deuterium discharge, a detector chip ($\#$00143) was processed using the same protocol as the five lowest energy \ce{^{241}Am} calibration detectors. Examples of tracks as they progressed with subsequent etchings are shown in \cref{fig:CR-39 00143 progressive etch tracks}. Table~\ref{tab:00143 progressive etch minor axes and incident energy for each etch interval} shows each etching interval's measured minor axis, along with the corresponding incident energy. The mean incident energy for all the etching intervals was $116 \pm 8.9$ keV. The minor axis versus etching time for the active chip and the calibration data are shown in \cref{fig:CR-39 00143 progressive etch minor axis vs etching time}. The slope for the active chip progressive etch was $2.46 \pm 0.06$ and using the slope versus energy calibration curve for \ce{^{241}Am} shown in \cref{fig:CR-39 - Minor axis vs energy minor axis vs etching time slope of minor axis versus energy}c corresponds to $106 \pm 3$ keV. 

\begin{table}
\centering
\caption{\label{tab:00143 progressive etch minor axes and incident energy for each etch interval}Minor axis and incident energy at each etch interval for 00143.}
\resizebox{0.35\columnwidth}{!}{%
\begin{tabular}{|c|c|c|}
\hline
\textbf{\begin{tabular}[c]{@{}c@{}}Etching Time\\ (hrs)\end{tabular}} & \textbf{\begin{tabular}[c]{@{}c@{}}Minor Axis\\ (um)\end{tabular}} & \textbf{\begin{tabular}[c]{@{}c@{}}Incident Energy \\ (keV)\end{tabular}} \\ \hline
1 & $8.81 \pm 0.88$ & $125 \pm 16.1$ \\ \hline
2 & $11.5 \pm 1.36$ & $91.9 \pm 20.8$ \\ \hline
3 & $13.8 \pm 1.72$ & $127 \pm 18.9$ \\ \hline
4 & $15.9 \pm 2.34$ & $99 \pm 24.2$ \\ \hline
5 & $18.8 \pm 2.48$ & $137 \pm 18.3$ \\ \hline
\end{tabular}%
}
\end{table}

The progressive etchings provide additional support that the incident particles were roughly 100 keV alphas. How a track progresses depends on the damage profile in the material, i.e., the amount of energy transferred from the penetrating particle that breaks bonds in the CR-39 polymer. The progression of a track cone is referred to as the track-etch rate and is covered in more detail in \cite{durraniSolidStateNuclear2013, fleischerNuclearTracksSolids1975,guoChapterSolidStateNuclear2012}. The plasma track-etch rates followed the same linear trend as the calibration tracks from \ce{^{241}Am}, which further supports a charged particle origin. 

\subsection*{\label{subsec:track depths}Plasma Track Depths}

\begin{figure}
  \centering
       \begin{subfigure}{0.3\textwidth}
        \includegraphics[width=1\textwidth]{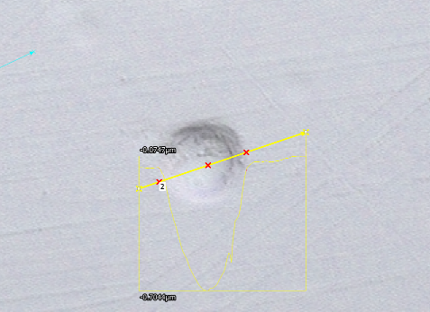}
        \caption{}
        \end{subfigure}
        \begin{subfigure}{0.3\textwidth}
        \includegraphics[width=1\textwidth]{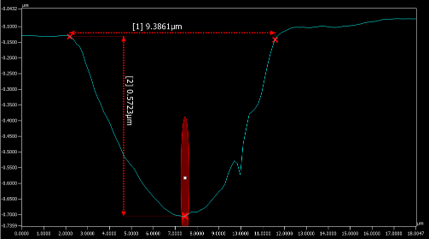}
        \caption{}
        \end{subfigure}
      \caption{Examples of tracks and their depth profiles taken using the Keyence VK-X1000.}
      \label{fig:CR-39 - example images of 00134 and 00132 track depth profiles}
\end{figure}

This section follows the same procedure to measure the track depth and minor axis diameter using the Keyence VK-X1000 3D laser scanning microscope as was done with the \ce{^{241}Am} calibration detectors. As with the \ce{^{241}Am} data, these measurements were done manually which limited the data to $\sim$200 tracks. The mean track depth was $0.40 \pm 0.17$ \textmu m with a mean minor axis diameter of $7.60 \pm 1.41$ \textmu m. The resulting incident energy from the diameter and depth measurements were $93 \pm 26$ keV and $99.8 \pm 17.3$ keV. Taking into account the attenuation effects of 5 cm of deuterium gas at 10 Torr, the diameter and depth measurements suggested that the emitted energies from the cathode were $109 \pm 29$ keV and $117 \pm 19$ keV, respectively. An example of a track depth profile is shown in \cref{fig:CR-39 - example images of 00134 and 00132 track depth profiles}.

As with the track progressions, the depth of the plasma tracks indicate $\sim$100 keV alpha particles. It seems unlikely that an alternative damage mechanism would produce tracks that followed the same track-etch rate and have the same depth profile as the proposed alpha particle. Further support for an energetic charged particle origin is presented in the following sections.

\subsection*{\label{subsec:trajectories}Incident Particle Trajectories}

The location of the tracks on the detector coupled with their depth profiles allows for an estimation of the trajectories of the incident particles that produced the tracks. A plasma discharge experiment was done over a range of pressures from 1 to 10 Torr and current densities from 5 to 60 mA/cm$^2$, to a fluence of $1.41 \times 10^{23}$ ions/cm$^2$, as this was in the initial stages of the project and the parameters that led to tracks were still unknown. It was this test that significantly increased confidence the shallow tracks were due to heavy charged particles impacting the detector. The procedure entailed using the 3D laser scanning microscope Keyence VK-X1000 to extract depth profiles and from those an angle of incidence was estimated. The angles of incidence in combination with the location of each measured track on the detector allowed for the representation of trajectories. 

\begin{figure}
\centering
\includegraphics[width=0.45\textwidth]{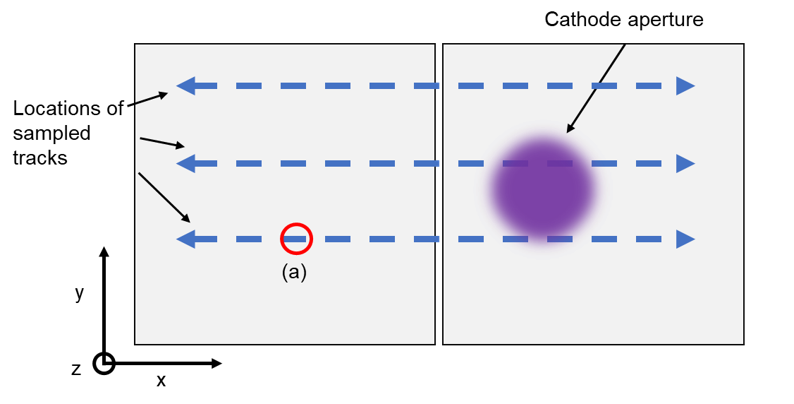}
\caption{Diagram of samples analyzed for track angles of incidence.}
\label{fig:CR-39 - diagram of 3 y-axis scanning locations}
\end{figure}

\begin{figure}
\centering
\begin{subfigure}{0.35\textwidth}
\includegraphics[width=1\textwidth]{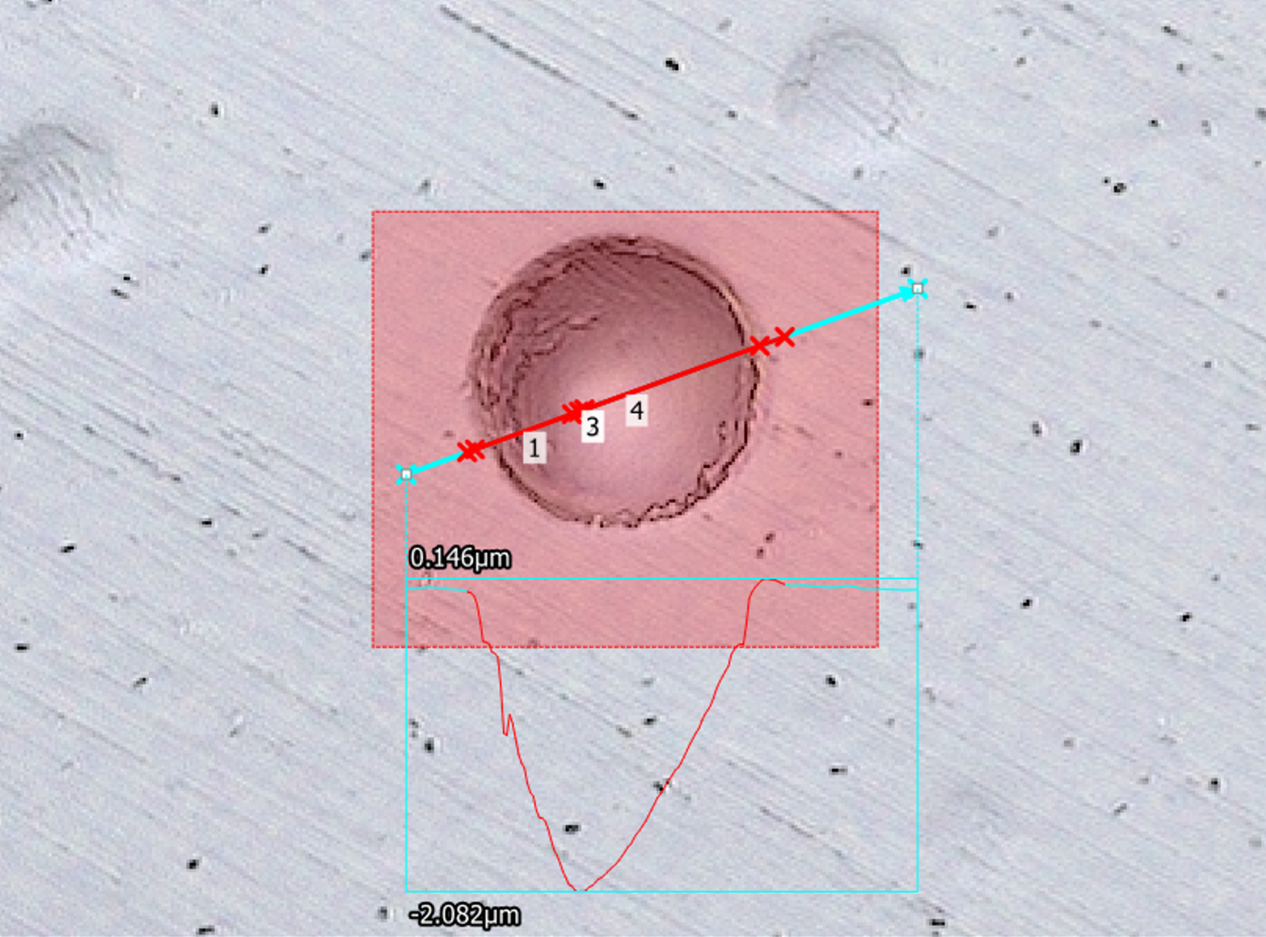}
\end{subfigure}
\begin{subfigure}{0.35\textwidth}
\includegraphics[width=1\textwidth]{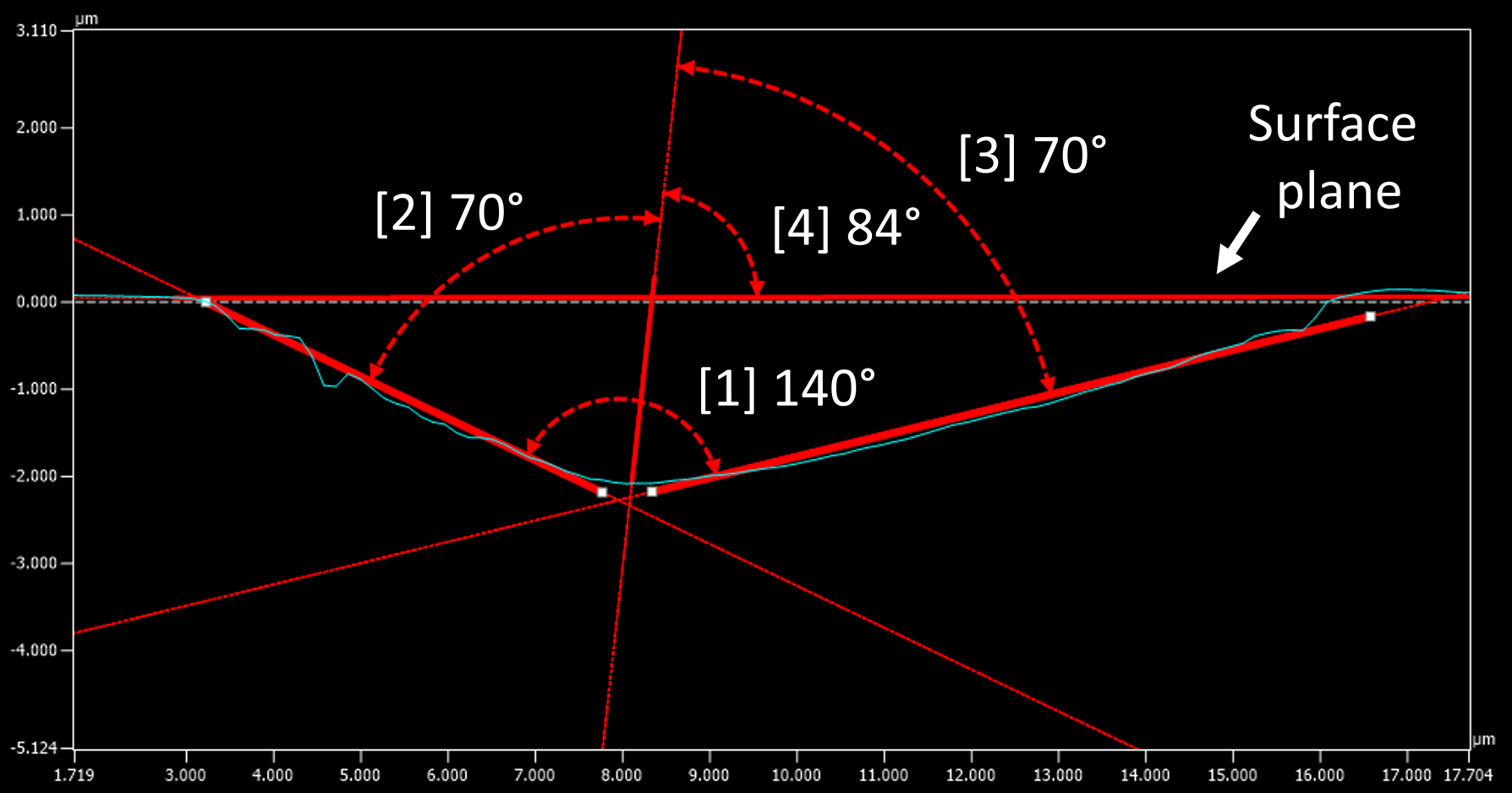}
\caption{}
\end{subfigure}
\caption{(a) Example of track used for trajectory estimate. (b) Estimating incident angles by bifurcating the track cone.}
\label{fig:CR-39 - example of track locations and angle of incidence}
\end{figure}

Two detectors were analyzed with the 3D microscope after the etching procedure. Three y-axis locations were scanned across each sample, as depicted in \cref{fig:CR-39 - diagram of 3 y-axis scanning locations}. An example of the track depth profile and angle of incidence estimate is shown in \cref{fig:CR-39 - example of track locations and angle of incidence}. The incident angles were estimated by first forming an angle between the track walls, angle [1] in \cref{fig:CR-39 - example of track locations and angle of incidence}b, which was $140 \degree$. Angle [1] was then bifurcated into two equal angles, [2] and [3], $70 \degree$. Then angle [4] is measured with respect to the surface of the CR-39 to produce the incidence angle which was $84 \degree$. A source of error results from the tracks being in the transition phase where they are not perfectly conical and have rounded sides which obfuscates the proper angle formed between the track walls. 

\begin{figure}
\centering
\begin{subfigure}{0.35\textwidth}
\includegraphics[width=1\textwidth]{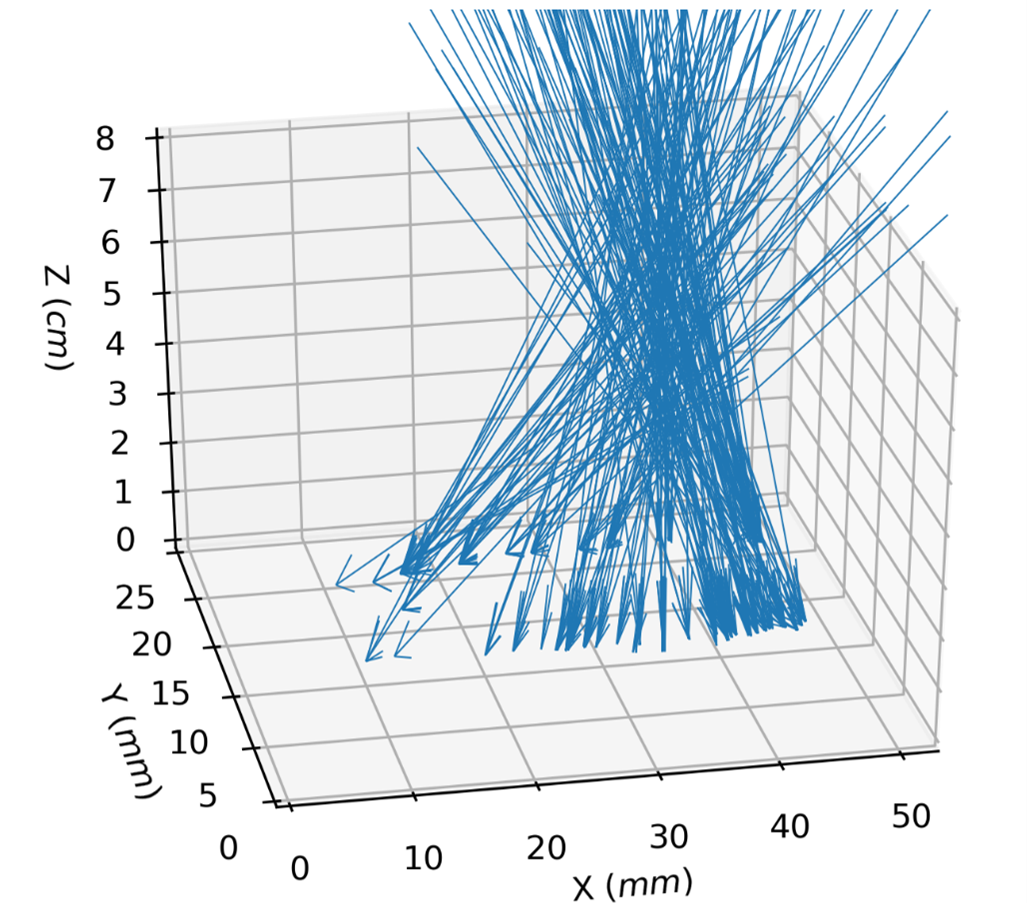}
\caption{}
\end{subfigure}
\begin{subfigure}{0.35\textwidth}
\includegraphics[width=1\textwidth]{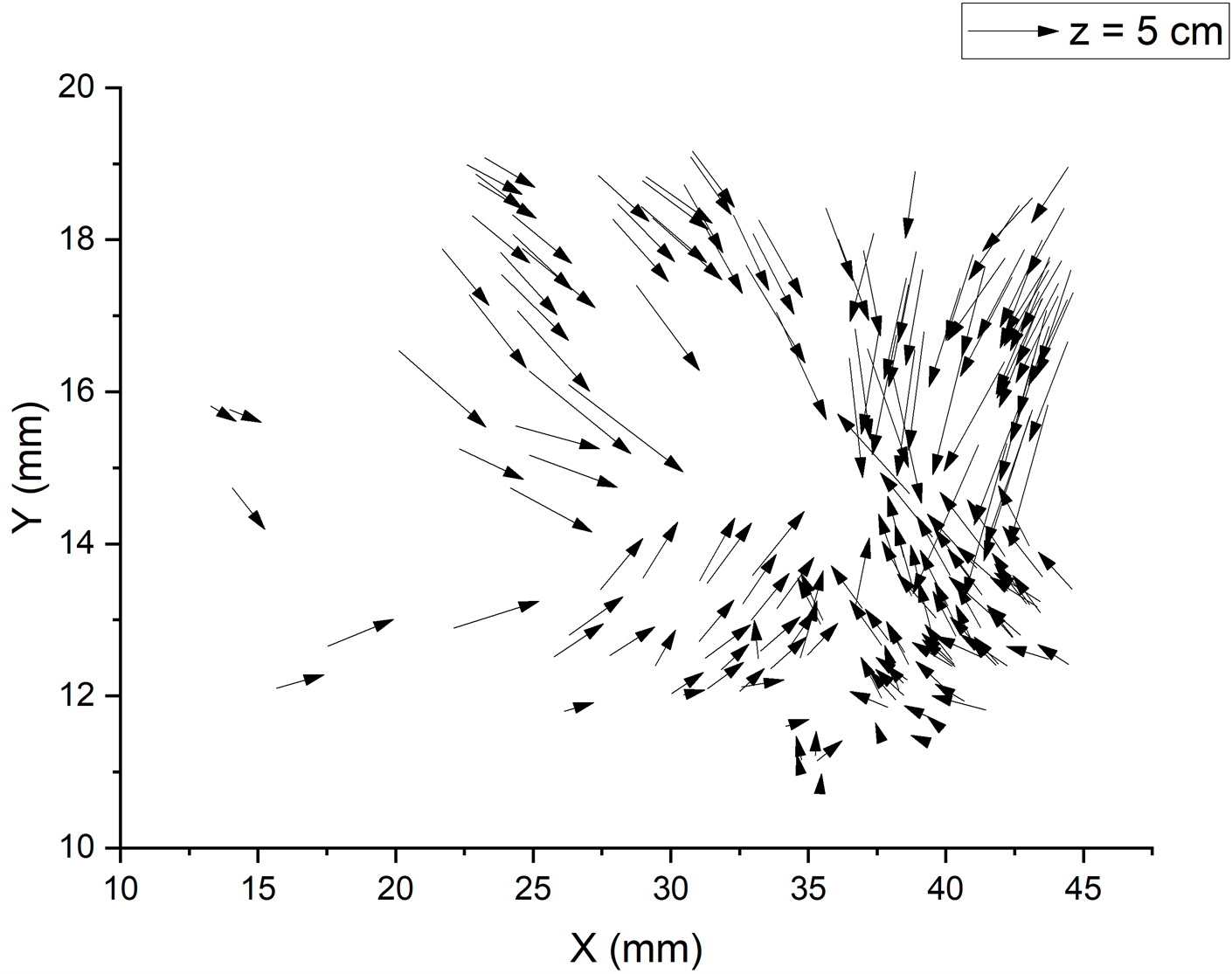}
\caption{}
\end{subfigure}
\caption{(a) 3D trajectories for particles that produced tracks in CR-39. (b) Cross-section of trajectories at z = 5 cm from the detector surface which is where the cathode is located.}
\label{fig:CR-39 - 3D trajectory plots and cross-section}
\end{figure}

The coordinates of the tracks, along with the angle of incidence, were then used to project the trajectories. An example of the trajectory plot is shown in \cref{fig:CR-39 - 3D trajectory plots and cross-section}a. The 2D projection of the trajectories for a cross-section at 5 cm from the detector surface is shown in \cref{fig:CR-39 - 3D trajectory plots and cross-section}b. It is apparent that the particles point toward a common point of origin at the cathode, which removes the possibility of cosmic rays coming from outside the chamber causing the tracks. The trajectories also remove the possibility that the particles originated from elsewhere in the chamber, e.g., the chamber walls. Hence, the environment, and subsequent mechanism, that produced the alphas is confined to the plasma and electrodes.

\subsection*{\label{subsec:Au on CR-39}Au Deposited on CR-39}

This section covers the deposition of gold on the CR-39 to investigate whether a type of charge buildup or plasma radical interaction caused the tracks in the plasma experiments. This test is also a method for energy spectroscopy. Depending on the thickness of gold, bounds are set on the minimum and maximum energies for an alpha particle to traverse the gold and retain enough energy to produce a track or be absorbed in the gold. An AJA ORION 3 sputter system was used to deposit 100 nm and 500 nm layers of gold on two detector chips. During the test, uncovered CR-39 was also present. The plasma parameters were 10 Torr D$_2$, 40 mA/cm$^2$, to a fluence of $5 \times 10^{21}$ ions/cm$^2$. A concern was that the gold layer would prevent the NaOH etching solution from interacting with the CR-39 and forming tracks. Preliminary Au deposited samples irradiated with unattenuated \ce{^{241}Am} were observed during the etching procedure and it was found that the gold layer quickly detached from the detector's surface once it was submerged in the $98 \degree$C 6.25 M NaOH and did not significantly hinder the etching process.

\begin{figure}
\centering
\begin{subfigure}{0.35\textwidth}
\includegraphics[width=1\textwidth]{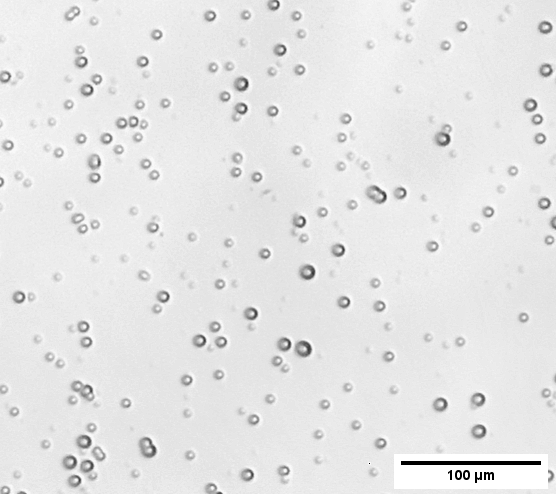}
\caption{}
\end{subfigure}
\begin{subfigure}{0.35\textwidth}
\includegraphics[width=1\textwidth]{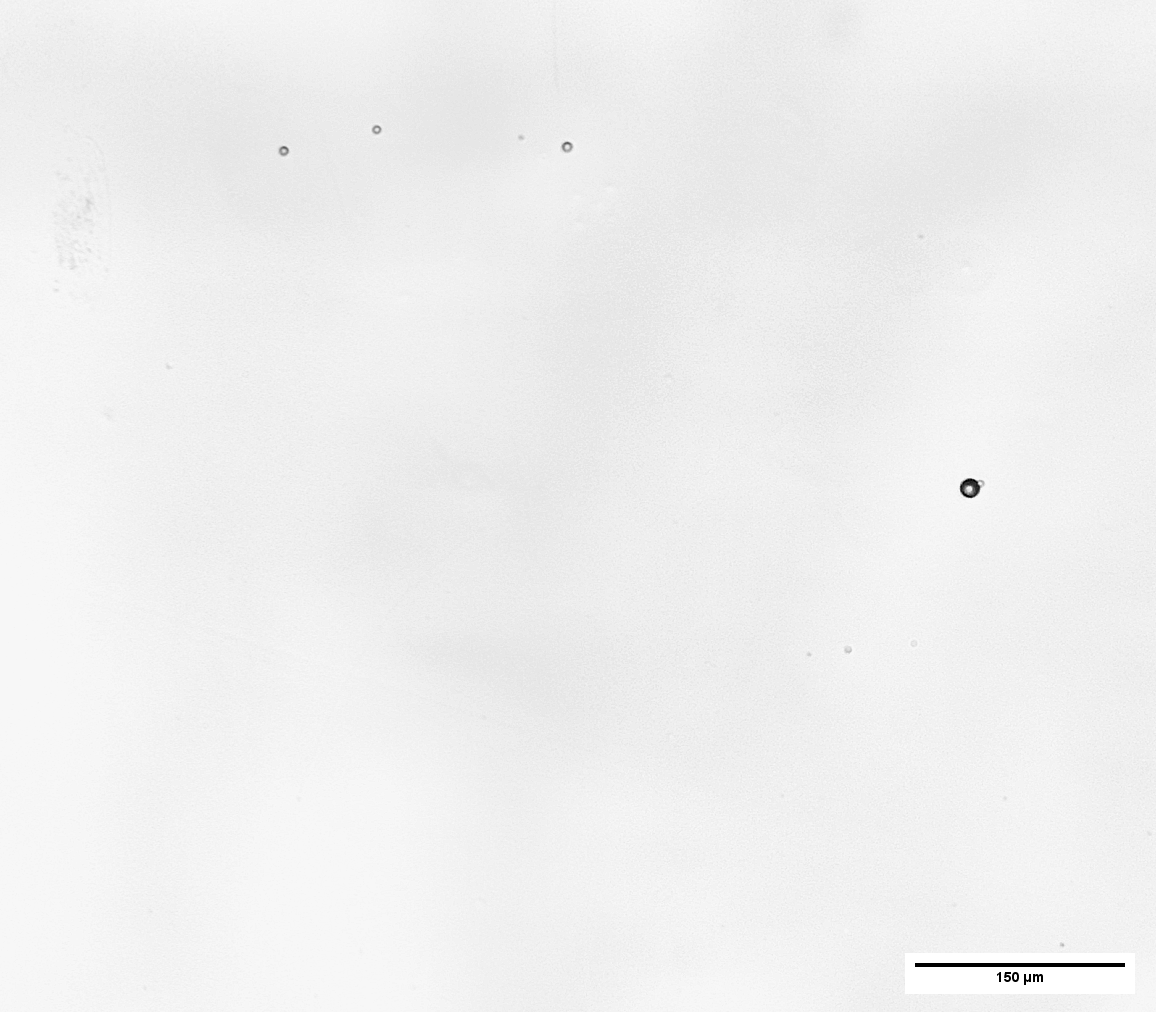}
\caption{}
\end{subfigure}
\caption{Tracks in the detector with (a) 100 nm Au and (b) 500 nm Au.}
\label{fig:CR-39 - tracks in 100 nm Au}
\end{figure}

Despite the 100 nm Au layer, elevated track densities were still present. The maximum track density was $3.52 \times 10^3$ tracks/mm$^2$/day. An example of tracks is shown in \cref{fig:CR-39 - tracks in 100 nm Au}a. The track minor axis from a bigaussian peak fit of the histogram was $7.6 \pm 0.6$ \textmu m. This corresponds to $102 \pm 11.1$ keV incident alphas on the CR-39 surface after transmission through the Au layer. The uncovered CR-39 chip had tracks with a minor axis of $8.57 \pm 1.09$ \textmu m which corresponds to $122 \pm 20$ keV incident alphas. A 122 keV alpha transmitted through 100 nm of gold has its energy attenuated to $88.8 \pm 6.9$ keV which is lower than the corresponding energies for the measured track minor axes. The depth of the gold layer was not measured, so it is possible less was deposited than desired. Or perhaps the Au deposition process leads to a type of sensitization of the CR-39 polymer that lowers the energy thresholds for track formation and hence leads to larger track sizes. Further studies are needed to determine how much of an impact the Au layer has on the etch rates. Regardless, the 100 nm Au layer removes the possibility that a charge build-up effect led to the tracks. The results also provide further evidence that a chemical reaction between plasma ions and the CR-39 does not cause the tracks. Supplemental evidence of this is from a test where the CR-39 was placed on the electrode holder, which produced no increased track densities.

\begin{figure}
\centering
\includegraphics[width=0.35\columnwidth]{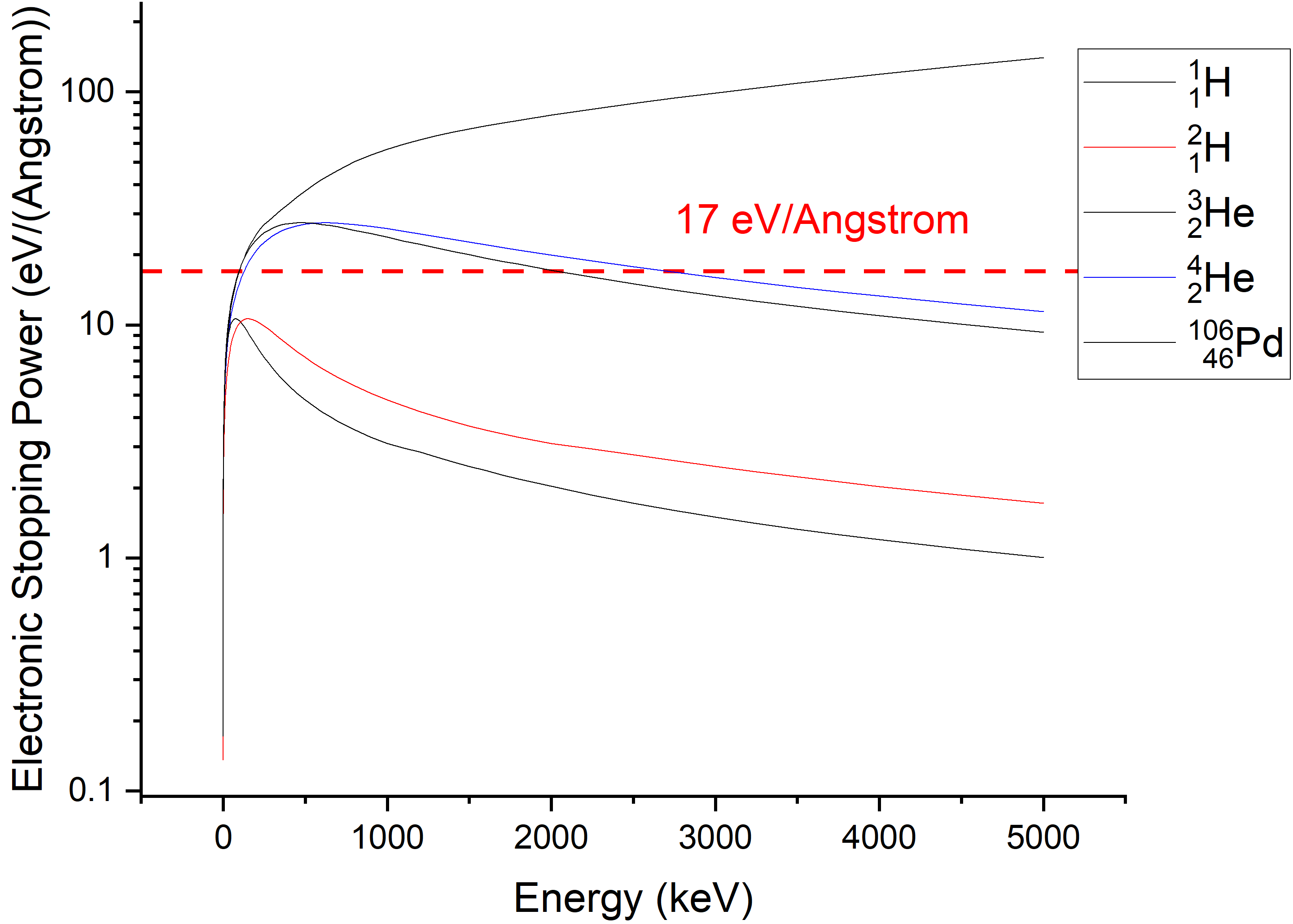}
\caption{\label{fig:TRIM - stopping power for protons alphas deuterons He3 Pd ions into CR39}Electronic stopping power of protons, deuterons, $^3$He, alphas, and Pd ions in CR-39 using a 0.957 compound correction.}
\end{figure}

A 122 keV incident alpha has a maximum electronic stopping power in CR-39 of 17 eV/\r{A}. The electronic stopping power in CR-39 for protons, alphas, deuterons, $^3$He$^+$, alphas, and Pd$^+$ ions are shown with a logarithmic y-axis in \cref{fig:TRIM - stopping power for protons alphas deuterons He3 Pd ions into CR39}. The data were produced using TRIM software (Transport of Ions in Matter) with a 0.957 compound correction \cite{zieglerSRIMStoppingRange2010}. Only $^3$He$^+$ and Pd$^+$ ions can produce a similar value in CR-39. Hence, protons and deuterons were removed from possible ions that produced the $8.57 \pm 1.09$ \textmu m diameter tracks. The electronic stopping power translates to the ionization of the medium by incident ions and recoils, which is the latent damage trail that produces tracks after etching \cite{fleischerNuclearTracksSolids1975, spohrIonTracksMicrotechnology1990,durraniSolidStateNuclear2013}. The 122 keV alpha then has a maximum ionization of 16.7 eV/\r{A} in CR-39, which is lower than the electronic stopping power, as energy loss due to phonons and high energy recoil electrons do not contribute to the latent damage trail. To obtain similar incident ion ionization values, $^3$He$^+$ requires an energy of 92 keV and 5.5 keV for Pd$^+$. Pd ions were included to investigate the possibility that the tracks were due to sputtered Pd from the electrodes. However, a 5.5 keV Pd$^+$ ion has an average range of only 26 \r{A} into Au. Therefore, even if there was a mechanism that accelerated Pd$^+$ ions to such energies, Pd$^+$ cannot be the species that produce the $7.6 \pm 6$ \textmu m diameter tracks in the CR-39 that was deposited with Au. 

With a 500 nm Au layer, the maximum track density dropped significantly to 63.5 tracks/mm$^2$/day. An example of the detector's surface with tracks still present is shown in \cref{fig:CR-39 - tracks in 100 nm Au}b. Using TRIM again, the maximum incident energy for an alpha particle to not produce a track with a 500 nm Au layer is 135 keV. The mean penetration distance of a 122 keV alpha into Au is $273 \pm 105$ \r{A}. Therefore, the 500 nm Au layer prevents alphas from reaching the CR-39 to produce a track. The maximum track density is within the range for the helium and hydrogen control tests, i.e., $78.5 \pm 37.3$ tracks/mm$^2$/day.

The Au deposited layer studies are consistent with $\sim$100 keV alpha particles incident on the detectors. The 100 nm Au tests not only showed that the energy transfer of the track producing particle is consistent with an alpha, but also removed multiple possibilities for what may have caused the tracks. First, static charge build-up on the surface of the dielectric CR-39 would no longer take place with the conductive Au layer, hence it is not what caused the tracks. Second, the gold layer prevents plasma radicals from interacting with the CR-39 surface and removes the possibility that the tracks were due to a plasma-material-interaction effect. Third, energetic Pd$^+$ ions would be unable to traverse the 100 nm Au layer which removes the possibility of sputtered Pd atoms from the electrodes producing the tracks. Last, the energy transfer of protons and deuterons is inconsistent with those particles being capable of producing tracks after attenuation through the gold. As for the 500 nm Au layer, it drastically reduced the track density from $3.52 \times 10^3$ tracks/mm$^2$/day for 100 nm Au to 63.5 tracks/mm$^2$/day. The reduction in track density also supports $\sim$100 keV incident alphas, as they are unable to penetrate the increased thickness. 

\section*{Discussion}

A novel automated analytical tool for CR-39 track detection was developed to investigate the emission of heavy energetic charged particles from a deuterium DC discharge with palladium electrodes. Extensive CR-39 detector experiments established the production of tracks that consistently corresponded to low energy alphas ($\sim$100 keV) emitted from the palladium electrodes. As helium is not present during the discharge, the emission of alpha particles indicates a nuclear process. However, the low energy alphas are in striking contrast to the work by Lipson et al. which used a pulsed deuterium discharge and CR-39 to monitor $dd$ fusion charged particle byproducts, e.g., 3 MeV protons and 1 MeV tritons \cite{Lipson2005h}. A reason Lipson et al.\cite{Lipson2005h} did not observe lower energy particle tracks is due to a 11 \textmu m thick Al foil placed in front of the CR-39, which would prevent the particles from reaching the detector. The Al foil was included to protect the detector from plasma interactions and sputtered electrode material. In the current work, the concern of plasma-CR-39 interaction leading to spurious results was addressed by placing the detectors farther from the discharge, tests with deposited gold on the detector surface, and comparisons with discharges using hydrogen and helium which showed no effect. 

It is possible a particle other than an alpha, but with similar stopping power in CR-39, $\sim$16.7 eV/\r{A}, was incident on the detectors. However, the gold deposition tests, progressive etchings, and track depth trends strongly suggest alpha emission. Further evidence that these tracks originated from the discharge electrodes came from using a Keyence VK-X1000 3D laser scanning microscope. The 3D microscopy technique allowed estimation of the incident particle trajectories. It was found that the tracks converged in proximity to the electrode holder and not from outside the chamber, which excludes cosmic rays or atmospheric radioisotopes (radon) as the source of the observed tracks, and limits the reaction site to the electrodes.

Some future research avenues are now briefly discussed. First, it is desirable to redo CR-39 calibrations while the chips are within a hydrogen environment similar to the experiments to account for possible hydrogen sensitization of the CR-39. To that effect, a differentially pumped duoplasmatron ion source is under construction to calibrate CR-39 with helium ions below 100 keV. Characterizing the ion source with a light-insensitive Passivated Implanted Planar Silicon (PIPS) detector then allows for more precise and accurate knowledge of the ion energy distributions incident on the CR-39. Next, the new calibration chips will be processed in an improved etching apparatus with $\pm$0.1$\degree$C temperature stability to reduce variation in track diameters and increase the energy spectrometry resolution. 

Work progressed for some time on implementing a diamond detector to complement the CR-39 data with prompt particle spectroscopy. However, the geometry and pressure of the system often leads to an erratic discharge and significant Electromagnetic Interference (EMI) prevented the diamond detector's use within the chamber. Characterizing the high-frequency I-V characteristics of the discharge then allows for the design of EMI filters to implement the prompt particle detectors and to determine whether components of the EMI are correlated to the alpha emission. 

Optical Emission Spectroscopy (OES) is also underway for a non-invasive plasma diagnostic technique. The use of a Langmuir probe was undesirable due to the small gap distance of the electrodes making it likely the probe would influence the discharge characteristics leading to erroneous conclusions. OES would then allow for the investigation of long lived metastable states that might lead to a type of energy step-up process for such high kinetic energy particles to exist in a low voltage discharge. Regardless of this conjecture, it seems natural to investigate the emission of other forms of low energy radiation such as soft x-rays as done by Miley et al.\cite{Miley2003} during DC glow discharges with Pd and Ti cathodes or low energy beta emission similar to fracto-emission\cite{dickinsonEmissionElectronsPositive1981}. Once further diagnostics are included, relationships between alpha production and system variables will be investigated, such as current density, high-frequency perturbations of the voltage, pressure, gap distance, gas mixtures, other electrode materials, and the effects of pre-discharge hydride-cycling of the electrodes for increased initial defect concentrations. Lastly, a subsequent paper will detail the cathode surface morphology and x-ray diffraction for the study of palladium-deuteride phases present after the deuterium discharge.

\bibliography{references}

\section*{Acknowledgements}

The authors acknowledge the use of facilities and instrumentation at the Materials Research Laboratory Central Research Facilities, University of Illinois, partially supported by NSF through the University of Illinois Materials Research Science and Engineering Center DMR-2309037. 

\section*{Author contributions statement}

E.P.Z. and G.H.M. conceived of the experiments. E.P.Z. then built the apparatuses described, conducted the experiments, and analyzed the results. The author's then discussed results. All authors reviewed the manuscript. 

\section*{Additional information}

\textbf{Competing interests}
This work was partially funded by NPL Associates, Inc and Industrial Heat, LLC. Neither organization directed the research or the interpretation of results.

\end{document}